%% file: paper.tex
\documentclass{sig-alternate}

\usepackage{def}
\usepackage{package}

\title{IOCA: High-Speed I/O-Aware LLC Management \\for Network-Centric Multi-Tenant Platform\vspace{-5ex}}
\date{}

\author{
{Yifan Yuan}\\
UIUC    
\and
{Mohammad Alian}\\
UIUC
 \and
 {Yipeng Wang}\\
Intel Labs
 \and
 {Ilia Kurakin}\\
Intel Corporation
 \and
 {Ren Wang}\\
Intel Labs
 \and
 {Charlie Tai}\\
Intel Labs
 \and
 {Nam Sung Kim}\\
UIUC
} %

\begin{document}
\maketitle

\input{abstract}

\input{introduction} 
\input{background}
\input{motivation}

\input{design}

\input{implementation}
\input{evaluation}

\input{discussion}
\input{related}
\input{conclusion}

\clearpage 


\bibliographystyle{IEEEtranS}
\bibliography{references}

\end{document}

%% file: abstract.tex
\begin{abstract}
\footnote{This work has been accepted by a conference. The authoritative version of this work "Don't Forget the I/O When Allocating Your LLC" will appear in the Proceedings of the 48th IEEE/ACM International Symposium on Computer Architecture (ISCA-48), 2021.}In modern server CPUs, last-level cache (LLC) is a critical hardware resource that exerts significant influence on the performance of the workloads, and how to manage LLC is a key to the performance isolation and QoS in the cloud with multi-tenancy. 
In this paper, we argue that besides CPU cores, high-speed network I/O is also important for LLC management. 
This is because of an Intel architectural innovation -- Data Direct I/O (DDIO) -- that directly injects the inbound I/O traffic to (part of) the LLC instead of the main memory. 
We summarize two problems caused by DDIO and show that 
(1) the default DDIO configuration may not always achieve optimal performance,
(2) DDIO can decrease the performance of non-I/O workloads which share LLC with it by as high as 32\%.

We then present \arch, the first LLC management mechanism for network-centric platforms that treats the I/O as the first-class citizen. 
\arch monitors and analyzes the performance of the cores, LLC, and DDIO using CPU's hardware performance counters, and adaptively adjusts the number of LLC ways for DDIO or the tenants that demand more LLC capacity. 
In addition, \arch dynamically chooses the tenants that share its LLC resource with DDIO, to minimize the performance interference by both the tenants and the I/O. 
Our experiments with multiple microbenchmarks and real-world applications in two major end-host network models demonstrate that \arch can effectively reduce the performance degradation caused by DDIO, with minimal overhead.
\end{abstract} 

%% file: introduction.tex
\section{Introduction}
\label{sec:intro}
The world has seen the dominance of Infrastructure-as-a-Service (IaaS)  in cloud data centers~\cite{moreno2012iaas}. 
IaaS hides the underlying hardware from the upper-level tenants and allows multiple tenants to share the same physical platform with virtualization technologies such as virtual machine (VM) and container (\ie, workload collocation). 
This not only facilitates the operation and management of the cloud but also achieves high efficiency and hardware utilization. 

However, the benefits of the workload collocation in the multi-tenant cloud do not come for free. 
Different tenants may contend with each other for the shared hardware resources, which often incurs severe performance interference~\cite{Nathuji:2010:QMP:1755913.1755938,Kambadur:2012:MIL:2388996.2389066,Govindan:2011:CQE:2038916.2038938,delimitrou2013ibench}. 
Hence, we need to carefully allocate and isolate hardware resources for tenants. 
Among these resources, the CPU's last-level cache (LLC), with much higher access speed than the DRAM-based memory and limited capacity (\eg, tens of MB), is a critical one~\cite{durner2019towards, Govindan:2011:CQE:2038916.2038938,Zhu:2016:DEQ:2872362.2872394,179040}.

There have been a large body of work on how to partition LLC for different CPU cores (and thus tenants) with hardware or software methods~\cite{qureshi2006utility,Albonesi:1999:SCW:320080.320119,Qureshi:2007:AIP:1250662.1250709,Balasubramonian:2000:MHR:360128.360153,Varadarajan:2006:MCC:1194816.1194856,Sanchez:2011:VSE:2000064.2000073,Bugnion:1996:CPC:237090.237195,Zhang:2009:TPP:1519065.1519076,ye2014coloris,rldrm,10.1145/2897937.2898036,7056026,10.1145/2872362.2872382,7920819}. 
For instance, recent Intel Resource Director Technology (RDT) enables LLC partitioning and monitoring on commodity hardware in cache way granularity~\cite{herdrich2016cache}. 
This spurs the innovation of LLC management mechanisms in the real world for multi-tenancy and workload collocation~\cite{selfa2017application,196286,el2018kpart,xu2018dcat,xiang2018dcaps,park2019copart}.
However, they all did not consider the role and impact of high-speed I/O in their works by Intel's Data Direct I/O (DDIO) technology~\cite{ddio}. 

Traditionally, inbound data from (PCIe-based) I/O devices is delivered to the main memory, and the CPU core will fetch and process it later. 
However, such a scheme is inefficient as to data access latency and memory bandwidth consumption. Especially with the advent of I/O devices with extremely high bandwidth (\eg, 100Gb network device~\cite{chakravarty2016100g} and 1Tb in the near future~\cite{eth-roadmap}) -- to the memory, CPU is not able to process all inbound traffic in time. As a result, Rx/Tx buffers will get overflown, and packet loss occurs. 
DDIO, instead, directly steers the inbound data to (part of) the LLC, and thus significantly relieves the burden of the memory (see \secref{sec:ddio}), which results in low processing latency and high throughput from the core.
In other words, DDIO lets the I/O share LLC's ownership with the core (\ie, I/O can also read/write cachelines), which is especially meaningful for network-centric, I/O-intensive platforms.

Typically, DDIO is completely transparent to the OS and applications. 
However, this may lead to sub-optimal performance, since (1) the network traffic fluctuates over time, and so does the workload of each tenant, and (2) I/O devices can contend LLC resource with the cores. 
Previously, researchers~\cite{tootoonchian2018resq,ousterhout2019shenango,neugebauer2018understanding,10.1145/3357223.3362737} have identified the ``\textbf{Leaky DMA}'' problem, \ie, the size of the NIC Rx ring buffer can exceed the LLC capacity for DDIO, which makes data move back and forth between the LLC and main memory. 
While ResQ~\cite{tootoonchian2018resq} proposed a simple solution for this by properly sizing the Rx buffer, our experiment shows that it often undesirably impacts the performance (see \secref{sec:ld}). 
On the other hand, we also identify another DDIO-related inefficiency, the ``\textbf{Latent Contender}'' problem (see \secref{sec:lc}). 
That is, without being aware of DDIO, the CPU core is assigned with the same LLC ways that the DDIO is using, which incurs inefficient LLC utilization. 
Our experiment shows that this problem can incur $32\%$ performance degradation even for non-networking workloads. 
These two problems indicate the deficiency of pure core-oriented LLC management mechanisms and necessitate the configurability and awareness of DDIO for extreme I/O performance.  

To this end, we propose \arch, the first, to the best of our knowledge, I/O-aware LLC management mechanism.
\arch periodically collects statistics of the core, LLC, and I/O activities by using CPU's hardware performance counters. 
Based on the statistics, \arch determines the current system state with a finite state machine (FSM), and then allocates the LLC ways for either CPU cores or DDIO adaptively. This helps alleviate the impact of the Leaky DMA problem.
In addition, \arch shuffles the LLC ways allocation to avoid memory-intensive tenants sharing the LLC ways with DDIO, 
so that the performance interference between the core and I/O (\ie, the Latent Contender problem) can be reduced.

We develop \arch as a user-space daemon in Linux and evaluate it on a commodity server system with multiple real-world applications and two popular end-host network models. 
Our results show that compared to a case running a single workload, applying \arch in co-running scenarios can restrict the performance degradation of both networking and non-networking applications to less than 10\%, while without \arch, such degradation can be as high as$\sim30\%$.

Overall, we make the following contributions in this paper: 

\begin{itemize}[leftmargin=*]
    \item We identify and summarize the inefficiencies of the current DDIO usages that cause sub-optimal performance in multi-tenant cloud environments, as well as the deficiency of the previous solution.
    \item We propose \arch, the first I/O aware LLC management mechanism, to achieve better performance isolation.
    \item We develop and evaluate \arch on a commodity server and show that it can better utilize DDIO, effectively reduce the performance interference between networking and non-networking applications to a relatively low level, compared to LLC partitioning that is unaware of DDIO.
\end{itemize}

%% file: background.tex
\section{Background}
\label{sec:back}

\begin{figure}[!t]
    \centering 
    \includegraphics[width=0.75\linewidth]{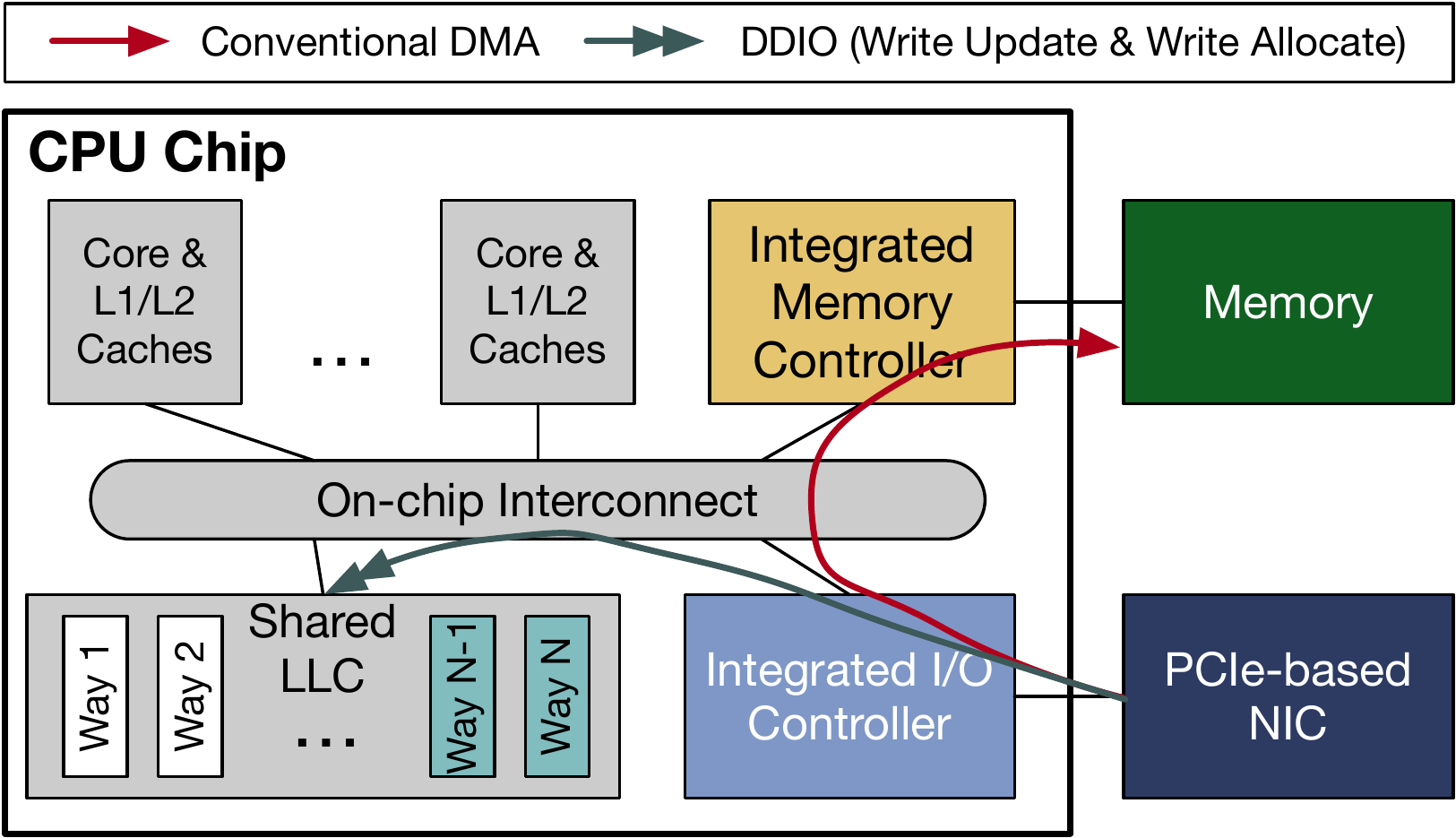}
    \caption{Typical cache organization in modern server CPU, conventional DMA path, and DDIO Technology for NIC.}
    \vspace{-3ex}
    \label{fig:ddio}
\end{figure}
\subsection{Managing LLC in Modern Server CPU}
In modern server-class CPUs, shared Last-Level Cache (LLC) is an important on-chip resource to achieve high performance. 
As studied in multiple works ~\cite{Govindan:2011:CQE:2038916.2038938,Kasture:2014:UEC:2541940.2541944,Mars:2011:BIU:2155620.2155650}, sharing LLC can cause performance interference among the collocated VM/containers. 
This motivates the practice of LLC monitoring and partitioning on modern server CPUs. 
Since the Xeon E5 v3 generation, Intel began to provide a set of techniques, named Resource Director Technology (RDT)~\cite{rdt}, for resource management in the memory hierarchy.
In this set, Cache Monitoring Technology (CMT) provides the ability to monitor the LLC utilization by different cores; Cache Allocation Technology (CAT) can assign LLC ways to different cores (and thus different applications or VMs/containers)~\cite{herdrich2016cache}.\footnote{Note that with CAT, a core can only allocate cachelines to its assigned LLC ways, but can still load/update cachelines from all the LLC ways.}
Programmers can leverage these techniques by simply accessing corresponding Model-Specific Registers (MSRs) or using high-level libraries~\cite{pqos}.
Furthermore, different mechanisms can be built atop RDT, to achieve dynamic management of LLC~\cite{selfa2017application,196286,el2018kpart,xu2018dcat,xiang2018dcaps,park2019copart,rldrm}.

\subsection{Data Direct I/O Technology}
\label{sec:ddio}

Conventionally, direct memory access (DMA) operations from a PCIe device\footnote{In this paper, we mostly focus on PCIe-based NIC.} use memory as the destination. 
That is, when transferring packets from the NIC to the host, the packet data will be written to the memory with addresses designated by the packet descriptors, as demonstrated in \figref{fig:ddio}. 
Later, when the CPU core has been informed about the ready packets, it will fetch the data from the memory to the cache hierarchy for future processing. 
However, as the bandwidth of the commodity network increases dramatically in the past decades, two drawbacks of such a DMA scheme become salient: 
(1) Accessing memory is relatively time-consuming, which can potentially limit the performance of packet processing. 
Suppose we have 100Gb inbound traffic. 
For a 64B packet with 20B Ethernet overhead, the packet arrival rate is 148.8 Mpps.
This means any component on the I/O path, like I/O controller or core, has to spend no more than 6.7ns on each packet, or packet loss will occur. 
(2) It consumes much memory bandwidth. 
Again with 100Gb inbound traffic, for each packet, it will be written to and read from memory at least once, which easily leads to $100Gb/s\times 2=25GB/s$ memory bandwidth consumption. 

To relieve the burden of memory, Intel proposed Direct Cache Access (DCA) technique~\cite{Huggahalli:2005:DCA:1069807.1069976}, which allows the NIC to write data directly to CPU's LLC. 
And in modern Intel Xeon CPUs, this has been implemented as Data Direct I/O Technology (DDIO)~\cite{ddio,ddio-brief}, which is transparent to the software.  
Specifically, as shown in \figref{fig:ddio}, when the CPU receives data from the NIC, an LLC lookup will be performed to check if the cacheline with the corresponding address is present with a valid state. 
If so, this cacheline will be updated with the inbound data (\ie, \texttt{write update}). 
If not, the inbound data will be allocated to the LLC (\ie, \texttt{write allocate}), and dirty cachelines may be evicted to the memory.
By default, DDIO can only perform \texttt{write allocate} on  two LLC ways (\ie, Way $N-1$ and Way $N$ in \figref{fig:ddio}). 
Similarly, with DDIO, a NIC can directly \texttt{read} data from the LLC; if the data is not present, the NIC will read it from the memory instead, but not allocate it in the LLC.
DDIO cuts the packet processing latency by removing the memory trip time~\cite{kurth_netcat:_2020}.
Also, with DDIO, the memory bandwidth consumption of the network packet processing can be significantly reduced~\cite{alian-ispass}. 

Note that although DDIO is Intel-specific, the basic concept is general, as other CPUs also have DDIO's counterparts, like ARM's Cache Stashing~\cite{stashing}.

\begin{figure}[t]
    \begin{subfigure}[b]{0.45\linewidth}
      \includegraphics[width=\linewidth]{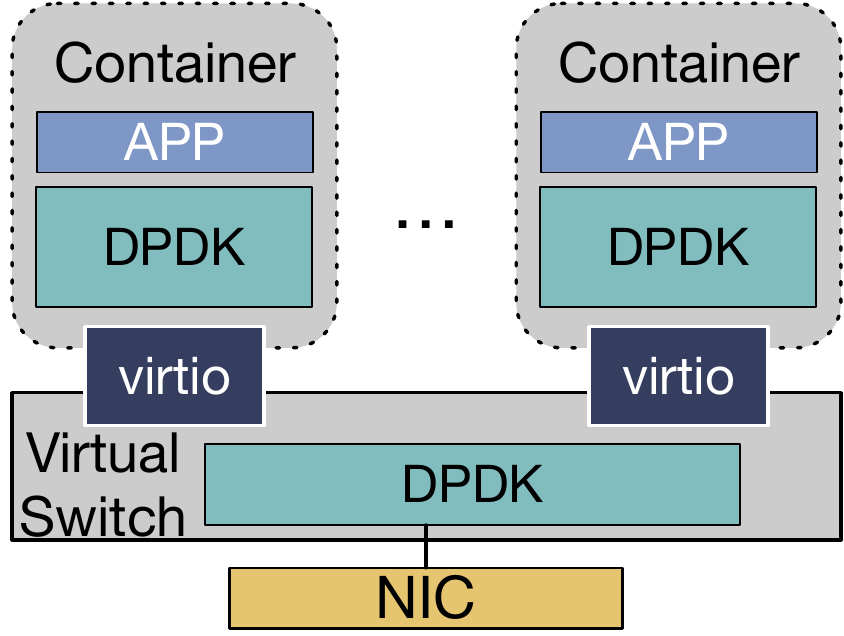}
        \caption{Aggregation.}
      \label{fig:agg}
    \end{subfigure}
     \hfill
    \begin{subfigure}[b]{0.45\linewidth}
      \includegraphics[width=\linewidth]{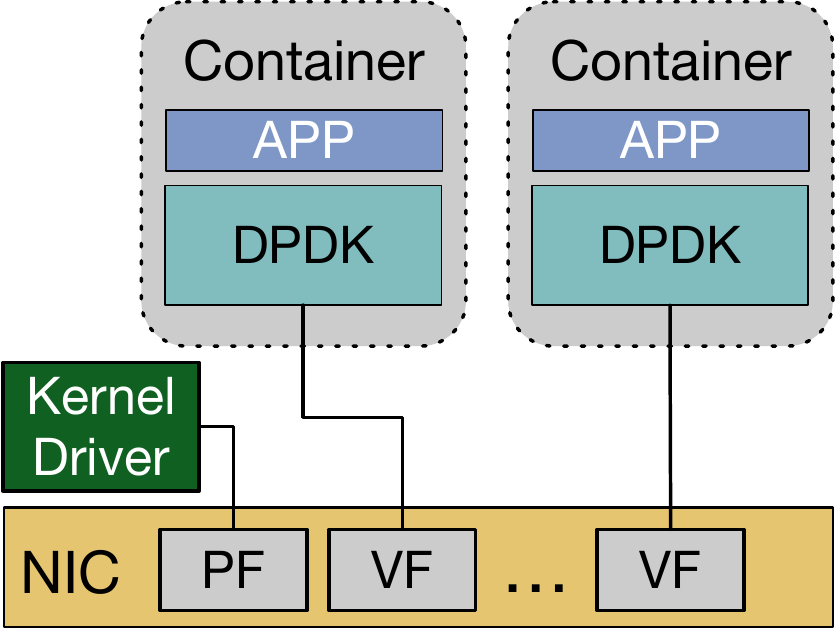}
      \caption{Slicing. }
      \label{fig:slicing}
    \end{subfigure}
     \vspace{-2ex}
     \caption{Two models of endhost network.}
      \label{fig:connect}
     \vspace{-2ex}
\end{figure}
\subsection{End-host Network in Virtualized Servers}
Servers in the modern cloud are usually virtualized. That is, multiple tenants (either VMs or containers) can co-exist in the same physical server by sharing hardware resources. 
Traditionally, these tenants are connected to the physical network and each other via a network bridge (\ie, Linux \texttt{bridge}).
As the trend of network virtualization~\cite{179731} and software-defined networking (SDN), however, the Linux bridges become unsatisfactory to the performance and functionality requirement of the end-host network. 
Hence, modern data centers adopt two models to tackle this problem (see \figref{fig:connect}). 
The key difference is the way they interact with the physical NIC.

First, SDN-compatible virtual switches, such as Open vSwitch (OVS)~\cite{ovs2} and VFP~\cite{smartnic2}, have been developed and deployed. 
As demonstrated in \figref{fig:agg}, the virtual switch controls the physical NIC and sends/receives packets to/from it. 
Tenants are connected to the virtual switch's bridges via interfaces like virtio~\cite{Russell:2008:VTD:1400097.1400108}.
With kernel-bypass network libraries (\eg, DPDK~\cite{dpdk}), such virtual switches can achieve high throughput and low latency. 
Since all traffic in this model needs to go through the virtual switch, we call this model ``aggregation''.

For the second model, the hardware-based single root input/output virtualization (SR-IOV) technique~\cite{dong2012high} has been leveraged. 
We depict this model in \figref{fig:slicing}. 
With SR-IOV, a single physical NIC can be virtualized to multiple virtual functions (VFs), each with dedicated hardware queue(s) ~\cite{vmdq}. 
While the physical function (PF) is still connected to the host OS/hypervisor, we can bind the VFs directly to the tenants, and thus bypass the host OS/hypervisor. 
In other words, the basic switching functionality is offloaded to the NIC hardware, and each tenant directly talks to the physical NICs for packet reception and transmission.
Since this model disaggregates the hardware resource and assigns it to different tenants, it is also called ``slicing''. 
Note that, many hardware offloading solutions for virtual switches~\cite{smartnic3,agilio} can be intrinsically treated as the slicing model.

%% file: motivation.tex

\begin{figure}[!t]
    \centering 
    \includegraphics[width=0.75\linewidth]{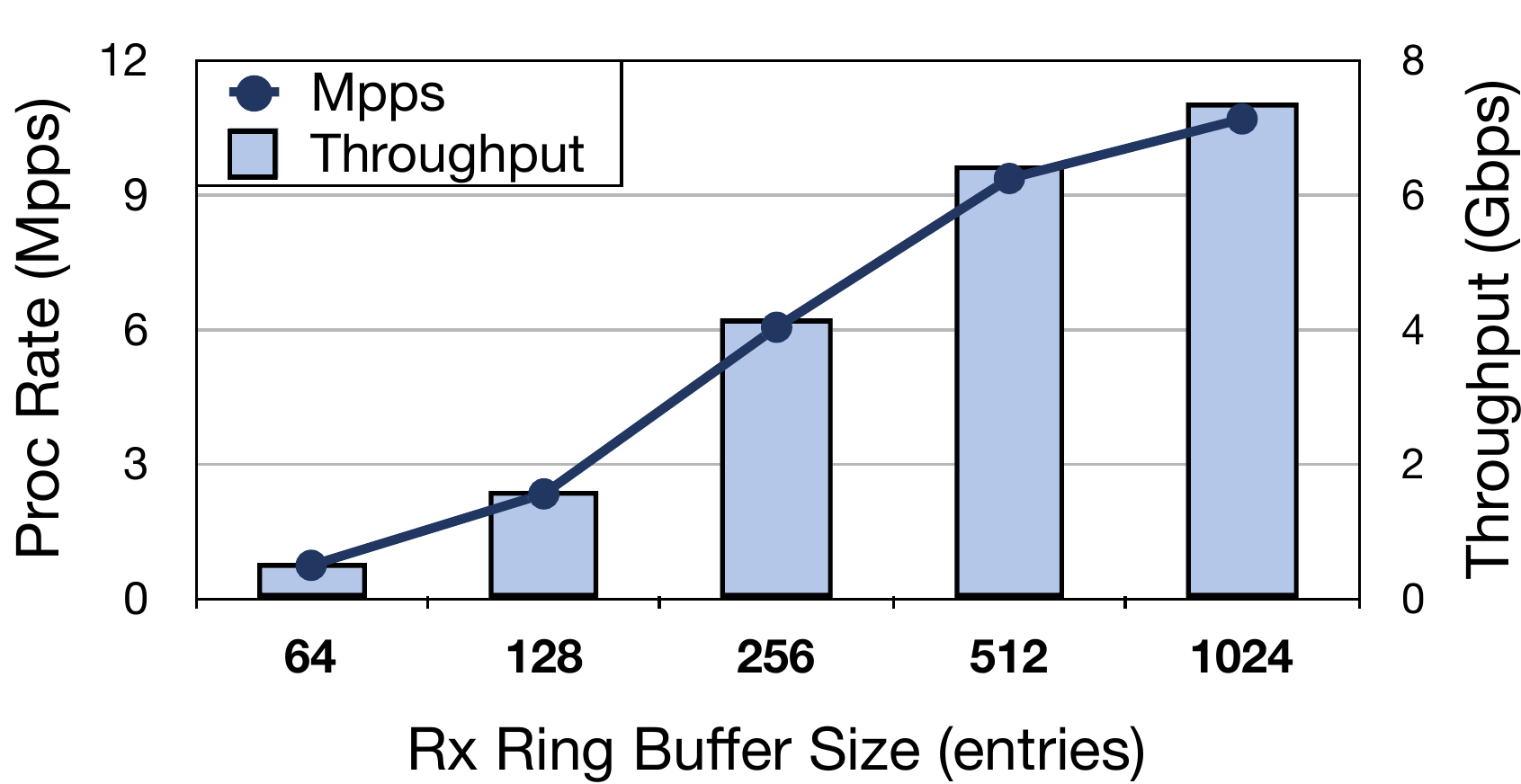}
    \vspace{-2ex}
    \caption{DPDK \textit{l3fwd} performance with different Rx ring buffer sizes in RFC2544 test.}
    \vspace{-3ex}
    \label{fig:data1}
\end{figure}
\section{The Impact of I/O on LLC}
\label{sec:mov}

In general, DDIO benefits the system performance. 
However, the static allocation of LLC does not always bring optimal performance due to two problems.
Here we discuss them separately. 

\subsection{The Leaky DMA Problem}
\label{sec:ld}
The first one is the ``Leaky DMA'' problem, as observed by multiple papers~\cite{tootoonchian2018resq,ousterhout2019shenango,neugebauer2018understanding,10.1145/3357223.3362737}. 
That is, since by default there are only two LLC ways for DDIO's \texttt{write allocate}, when the inbound data rate (\ie, NIC Rx rate) is higher than the rate that CPU cores can process, it is highly possible that the data in LLC waiting for processing is evicted to the memory by the newly incoming data, and later is brought back to the LLC again when cores need it. 
This incurs extra memory read/write bandwidth consumption, as well as increases the latency of processing each packet and eventually leads to a performance drop.  

In ResQ, the authors propose to solve this problem by reducing the size of the Rx/Tx buffers.
However, this workaround is not flawless. 
In a cloud environment, tens or even hundreds of VMs/containers can share a couple of physical ports through the virtualized network~\cite{netvm:nsdi14,container:sosp17}. 
If the total count of entries in all buffers is maintained below the default DDIO's LLC capacity, each VM/container only gets a very shallow buffer.
For example, assume that in the SR-IOV setup where we have 20 containers, each assigned a virtual function to receive traffic, and that each packet is around 1KB. 
To guarantee all the queues can be accommodated in the DDIO's cache capacity (several MB), each buffer can only have a small number of entries. 
A shallow Rx/Tx buffer can lead to severe packet drop issues, especially when we have bursty traffic, which is ubiquitous in modern cloud services~\cite{211225}.

Here we run a simple experiment to demonstrate such inefficiency (see \secref{sec:setup} for details of our testbed). 
We set up DPDK \textit{l3fwd} application to do TCP traffic routing. This application will look at the header of each network packet up against a flow table of 1 million flows. 
The packet is forwarded if a match is found. We run an RFC2544 non-packet-drop test~\cite{rfc2544} with the 40GbE NIC and a traffic generator machine.
From the results in \figref{fig:data1}, we see that by cutting half the buffer size (from 1024 to 512), the maximum throughput can drop by 13\%. 
If we use a small buffer of 64 entries, the throughput is less than one-tenth of the original throughput. 
This motivates us not to merely reduce the Rx/Tx buffer size, but to tune the DDIO's LLC capacity adaptively.

\begin{figure}[t]
    \centering
    \begin{subfigure}[b]{0.48\linewidth}
      \centering
      \includegraphics[width=\linewidth]{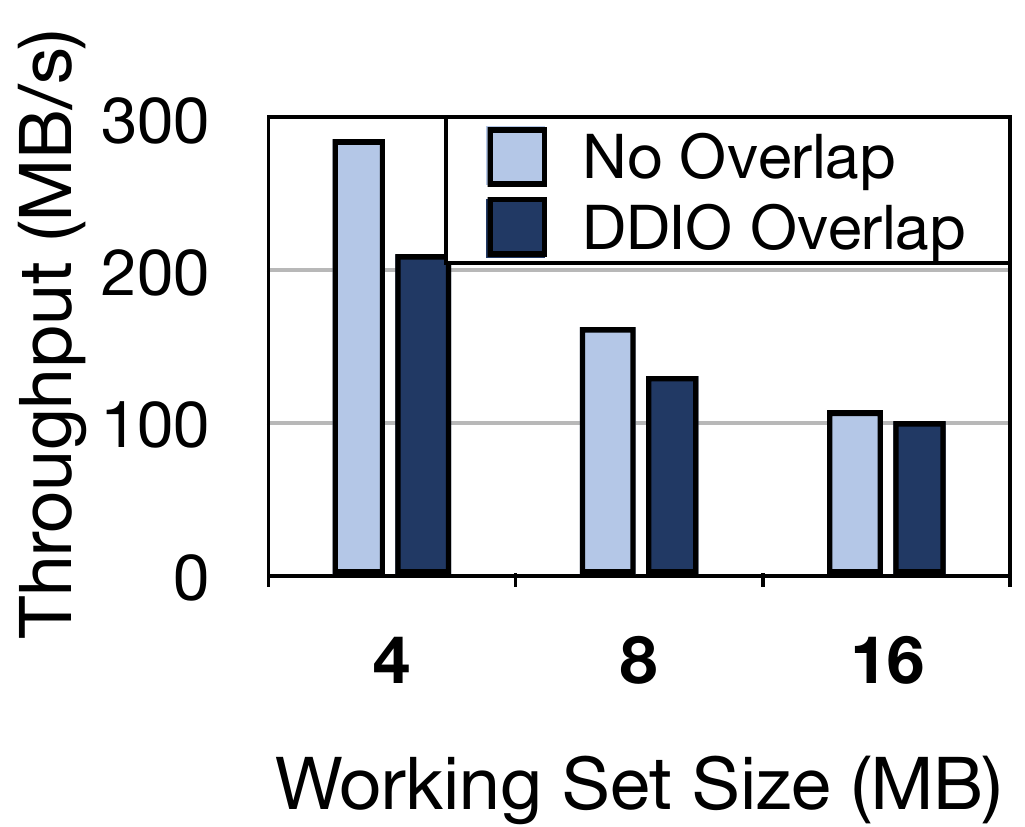}
        \caption{Throughput.}
      \label{fig:data2-1}
    \end{subfigure}
     \hfill
    \begin{subfigure}[b]{0.48\linewidth}
      \centering
      \includegraphics[width=\linewidth]{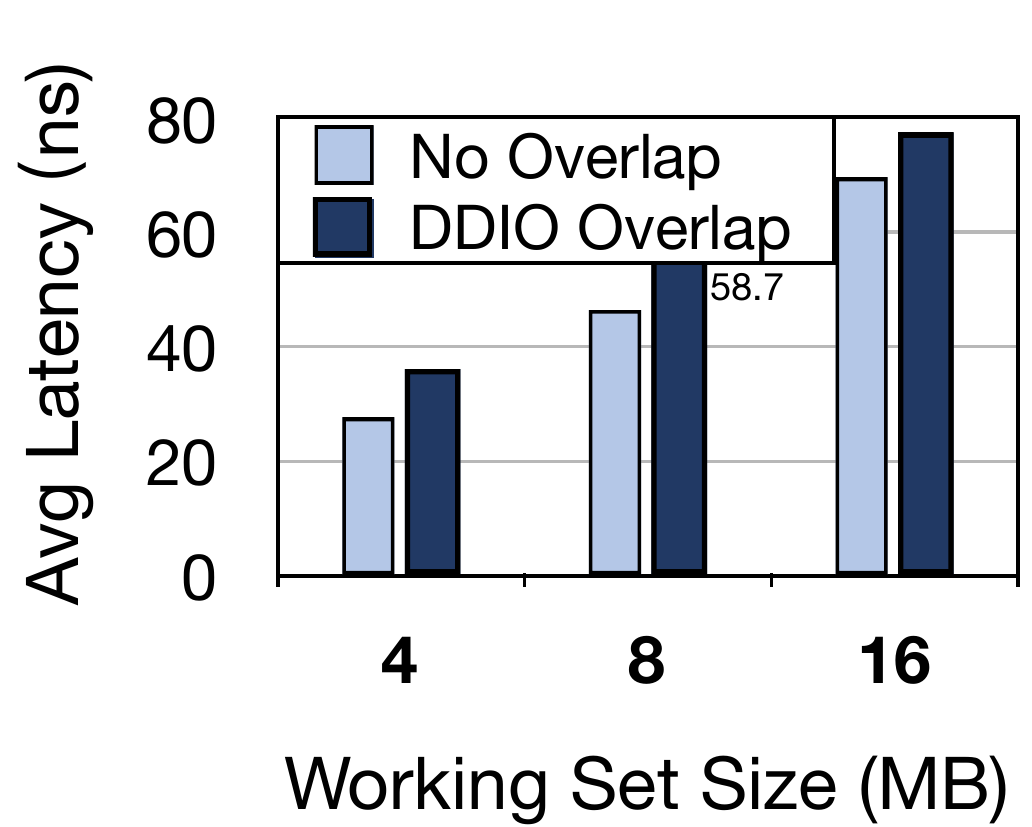}
      \caption{Latency.}
      \label{fig:data2-2}
    \end{subfigure}
     \vspace{-2ex}
     \caption{DDIO effect on X-Mem performance.}
      \label{fig:data2}
     \vspace{-4ex}
\end{figure}
\subsection{The Latent Contender Problem}
\label{sec:lc}
We identify a second problem caused by DDIO -- the ``Latent Contender'' problem. 
That is, since none of the current LLC management mechanisms is I/O-aware, when allocating LLC ways for different cores with CAT, they may unconsciously allocate DDIO's LLC ways to certain cores running LLC-sensitive workloads. 
This means even if, from the core point of view, these LLC ways are completely isolated, DDIO is actually still contending with the cores for the capacity.

We run another experiment to further demonstrate this problem. 
In this experiment, we first set up a container, which is bound to one CPU core, two LLC ways (\ie, Way $0-1$) and one NIC VF. 
This container is running DPDK \textit{l3fwd} with 40Gb traffic. 
We then set up another container, which is running on another core. 
We run X-Mem~\cite{gottscho2016x}, a microbenchmark for memory behavior of the cloud applications. 
We increment the working set of X-Mem from 4MB to 16MB\footnote{Since the LLC is non-inclusive, we need a working set larger than L2 cache size (1MB).} and apply the random-read memory access pattern.
We measure the average latency and throughput of X-Mem in two cases: (1) the container is bound to two dedicated LLC ways (\ie, no overlap), and (2) the container is bind to the two DDIO's LLC ways (\ie, DDIO overlap). 
As the results in \figref{fig:data2} show, even if X-Mem and \textit{l3fwd} do not explicitly share any LLC ways from the core point of view, DDIO may still worsen X-Mem's throughput by at most 26.0\% and average latency by at most 32.0\%. 
This lets us think of how we should select tenants that share LLC ways with DDIO.

%% file: design.tex
\section{IOCA Design}
\label{sec:design}

\begin{figure}[!t]
    \centering 
    \includegraphics[width=0.75\linewidth]{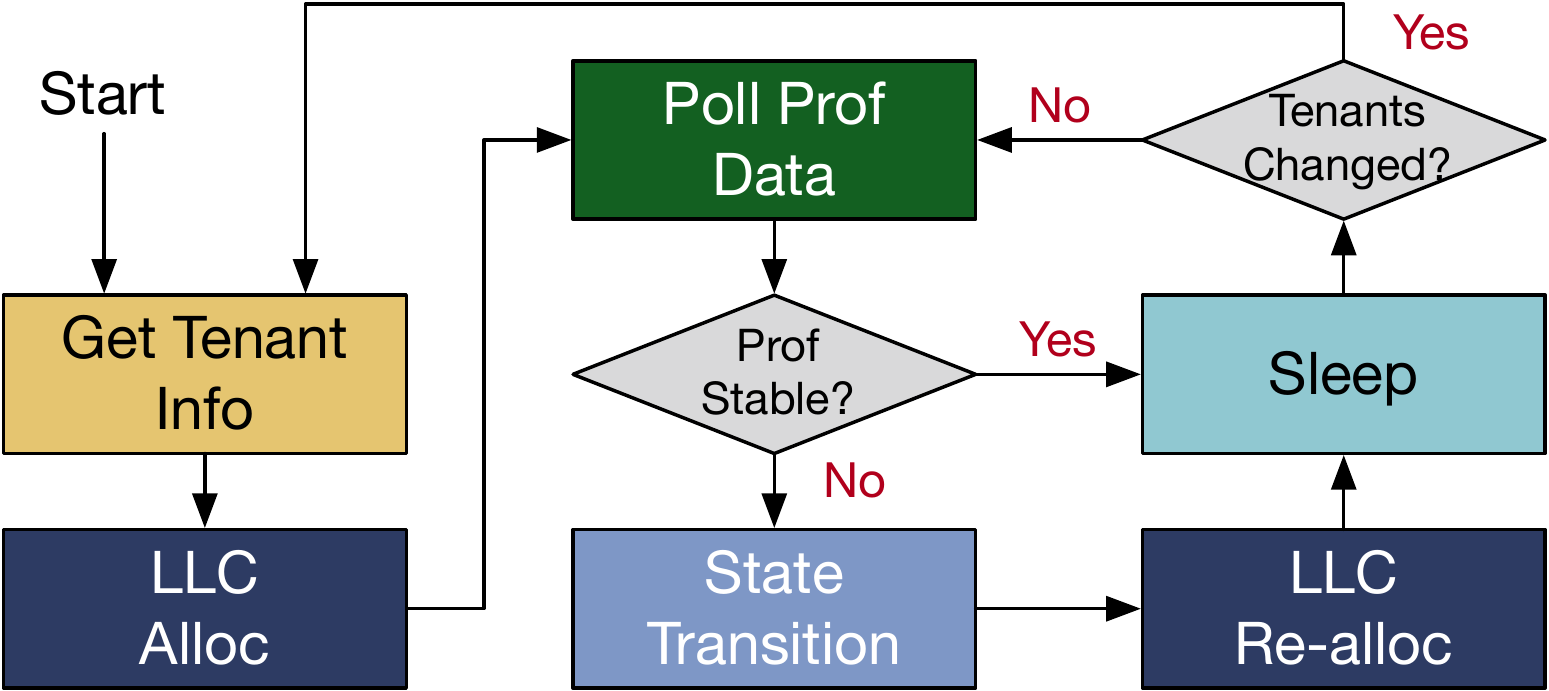}
    \vspace{-1ex}
    \caption{Execution flow of \arch.}
    \vspace{-3ex}
    \label{fig:flow}
\end{figure}

\arch is an I/O-aware LLC management mechanism that makes better use of DDIO technology for various situations in multi-tenant servers. 
We achieve this by dynamically configuring the LLC ways for DDIO, considering the characteristics of both the I/O and the core. 
When \arch detects an increasing amount of LLC misses from DDIO traffic, it first decides whether the misses are caused by the I/O traffic or by the application running in the cores.
Based on the decision, \arch allocates more or fewer LLC ways to either the core of the DDIO to mitigate the core-to-I/O or I/O-to-I/O interference.
\arch can also shuffle tenants' LLC allocation to further reduce core-I/O contention.
More specifically, \arch performs six steps to achieve its objective: \texttt{Get Tenant Info}, \texttt{LLC Alloc}, \texttt{Poll Prof Data}, \texttt{State Transition}, \texttt{LLC Re-alloc}, and \texttt{Sleep}. These steps are organized in a way demonstrated in \figref{fig:flow}. 
We describe each of the steps in detail in this section.

\subsection{Get Tenant Info and LLC Alloc}
At initialization (or tenants change), \arch obtains information of the tenants and the available hardware resources through the \texttt{Get Tenant Info} step.
For hardware resources, it needs to know the allocated cores and LLC ways for each tenant. 
For software, it requires two pieces of information. 
The first one is whether the tenant's workload is ``networking'' or not. 
This can help \arch to decide whether a performance fluctuation is caused by I/O or not, since non-I/O applications also have execution phases with different behaviors.
Note that a non-networking tenant may maintain the basic connection to the network (for ssh, \etc), but does not trigger intensive I/O traffic as a ``networking'' tenant.
The second one is the priority of each tenant. 
To improve resource utilization, modern data centers tend to collocate workloads with different priorities on the same physical platform~\cite{Mars:2011:BIU:2155620.2155650,Govindan:2011:CQE:2038916.2038938,Han:2016:IMD:2872362.2872388}.
Since the cluster management software commonly provides hints for such priorities~\cite{Verma:2015:LCM:2741948.2741964,10.1145/3342195.3387517}, \arch can obtain such information directly. 
In \arch,  we assume two possible priorities for each workload -- ``performance-sensitive'' (PS) and ``best-effort'' (BE)\footnote{BE does not necessarily mean a bandwidth-dominant application.} to each tenant. 
This is to demonstrate the basic idea and mechanism of \arch, and the types of priorities can be extended in real-world deployment.

\arch maintains a table to store (and update) aforementioned information for all the tenants.
Although the virtual switch is not a tenant, we still keep the record for it in the table and assign it with a special priority. 
After getting the tenant information, \arch allocates the LLC ways for each tenant accordingly (\ie, \texttt{LLC Alloc})\footnote{We assume the cores have been assigned to each tenant by the orchestrator such as Kubernetes.}.

\subsection{Poll Prof Data}
In this step, \arch polls the performance status of each tenant to decide the optimal LLC allocation.
Using the application-level metrics (operations per second, tail latency, \etc) is not a good strategy here since such metrics vary across tenants, and thus are difficult to be reasoned about. 
Instead, we directly get the profiling statistics of hardware events by polling the hardware performance counters. 
Specifically, we collect the statistics of the following events in \arch:

\niparagraph{Instruction per cycle (IPC).} IPC is a commonly-used metric to measure the execution performance of a program on a CPU core~\cite{bitirgen2008coordinated,Moreto:2009:FQF:1531793.1531806,Kambadur:2012:MIL:2388996.2389066,179040}. 
We also use it to detect tenants' performance degradation and improvement. 
IPC can be derived from \textit{retired\_instruction\_count / unhalted\_cycles}.

\niparagraph{LLC reference and miss.}
LLC reference and miss counts reflect the memory access characteristic of a workload. 
We can also derive the LLC miss rate from these values. 
LLC miss rate is yet another critical metric for workload performance~\cite{Blagodurov:2010:CSM:1880018.1880019,10.1145/2150976.2151021}.

\niparagraph{DDIO hit and miss.} DDIO hit is the number of DDIO transactions that apply \texttt{write update}, meaning the targeted cacheline is already in the LLC; DDIO miss reflects the number of DDIO transactions that apply \texttt{write allocate},  which indicates a victim cacheline has to be evicted out of the LLC for the allocation. 
These two metrics reflect the intensity of the I/O traffic, and the pressure it puts on the LLC. 

IPC and LLC reference/miss are per-core metrics, so we collect the data from each core that is occupied by tenants. 
If a tenant is occupying more than one core, we aggregate the values as the tenant's result.
DDIO hit/miss are chip-wise metrics, which means we only need to collect them once per CPU, and cannot distinguish between DDIO hit/miss caused by different devices or applications.

After collecting these events' data, \arch will compare them with the ones collected during the previous iteration. 
If the delta of one of the events is larger than a threshold \textit{THRESHOLD\_STABLE} (\ie, the profiling is not stable), \arch will jump to the \texttt{State Transition} step to determine how to (potentially) adjust the LLC allocation. 
Otherwise, it will regard the status of the system as unchanged and jump to \texttt{Sleep} step, waiting for the next iteration. 
However, there are three cases where we do not jump to the \texttt{State Transition} step: 
(1) If we only see IPC change, but no significant LLC reference/miss and DDIO hit/miss count change, we assume that this change is attributed to neither cache/memory nor I/O. 
(2) If we observe IPC change of a non-networking tenant (no DDIO overlap) with corresponding LLC reference/miss change, but no significant DDIO hit/miss count change over the system, we know this is mainly caused by CPU core's demand of LLC space. 
In this case, other existing mechanisms~\cite{selfa2017application,196286,el2018kpart,xu2018dcat,xiang2018dcaps,park2019copart} can be called to allocate LLC ways for the tenant.
(3)  If we observe IPC change of a non-networking tenant (with DDIO overlap) with corresponding LLC reference/miss change and DDIO hit/miss change, we will try shuffling LLC ways allocation (see \secref{sec:realloc}) at first.

\begin{figure}[!t]
    \centering 
    \includegraphics[width=0.7\linewidth]{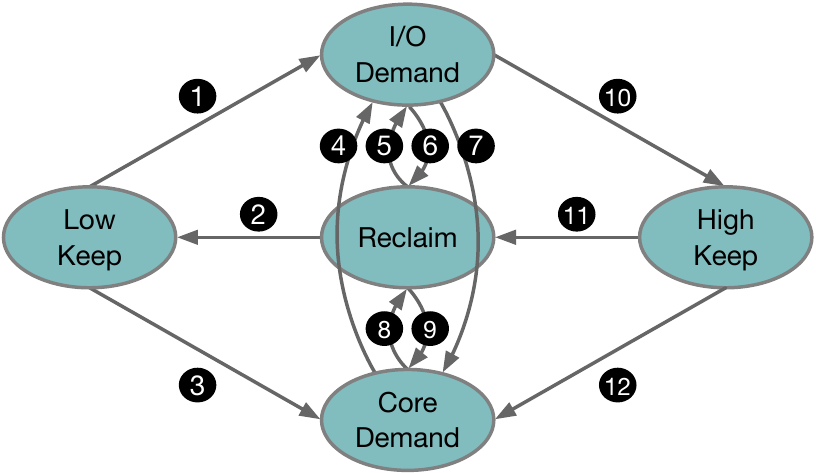}
    \vspace{-2ex}
    \caption{State transition diagram of \arch.}
    \vspace{-4ex}
    \label{fig:fsm}
\end{figure}
 
\subsection{State Transition}
The core of \arch design is a system-level (not per-tenant) FSM, which determines the current state of the system based on the profiling data from \texttt{Poll Prof Data}.
We depict the \arch FSM in \figref{fig:fsm}. In total, there are five states in the FSM: 

\niparagraph{Low Keep.}
In this state, the I/O traffic is not intensive and does not press the LLC (\ie, does not contend with cores for the LLC resource). 
\arch is in this state if the DDIO miss count is small.
Note that the value of DDIO hit count is not necessarily small, since if most DDIO transactions can end up with \texttt{write update}, which does not trash the LLC. 
Since I/O traffic does not trigger extensive cache misses, we always keep the number of LLC ways for DDIO at the smallest value (\textit{DDIO\_WAYS\_MIN}).

\niparagraph{High Keep.}
This is a state where we have already allocated the largest number of LLC ways for DDIO (\textit{DDIO\_WAYS\_MAX}), regardless of the numbers of DDIO miss and hit. 
We set such upper bound because we do not expect DDIO to compete with cores without any constraints across the entire LLC, especially when there is PS tenant running in the system with high priority.

\niparagraph{I/O Demand.}
This is a state where the I/O contends with cores for the LLC resource. In this state, I/O traffic becomes intensive, and the LLC space for \texttt{write update} can not satisfy the demand of DDIO transactions. 
As a result, more \texttt{write allocate} (DDIO miss) happen in the system, causing a large amount of cacheline evictions.

\niparagraph{Core Demand.}
In this state, the I/O also contends with cores for the LLC resource, but the reason is different. 
Specifically, now the core demands more LLC space. 
In other words, a memory-intensive networking application is running on the core. 
As a result, the Rx buffer is frequently evicted from the LLC ways that are allocated for the core, which leads to a decrease of the DDIO hit and an increase of the DDIO miss.

\niparagraph{Reclaim.}
Similar to Low Keep, in this state, the I/O traffic is not intensive either. 
The difference is, the number of LLC ways for DDIO is at a medium level, which is potentially wasteful. 
In this case, we should consider reclaiming some LLC ways from DDIO. 
Also, the LLC ways for a specific tenant can be more than enough, which motivates us to reclaim LLC ways from the core.

We then describe the transitions between states. \arch starts from the Low keep state. When the number of DDIO miss is greater than a threshold \textit{THRESHOLD\_MISS\_LOW}, it indicates that the current LLC ways for DDIO are insufficient. 
\arch determines the next state by further examining the value of DDIO hit and LLC reference. 
Decrease of the DDIO hit count with more LLC references implies the core(s) is increasingly contending the LLC with DDIO, and the entries in the Rx buffer(s) are frequently evicted from the LLC. 
In this case, \arch moves to the Core Demand state (\circled{3}). 
Otherwise, the DDIO miss is attributed to the more intensive I/O traffic. 
Hence, we move to the I/O Demand state (\circled{1}). 

In the Core Demand state, we regard the decrease of the DDIO miss rate as a signal of system balance and will go back to the Reclaim state (\circled{8}). 
If we observe an increase of DDIO miss count without fewer DDIO hits, we go to I/O Demand state (\circled{4}) since right now, the core is no longer the major competitor. 
If we observe neither of the two events, \arch will stay at the Core Demand state. 

In the I/O Demand state, if we still observe a large amount of DDIO miss, we keep in this state until we have allocated \textit{DDIO\_WAYS\_MAX} number of ways to DDIO, and then transit to High Keep state (\circled{10}). 
If a significant degradation of DDIO miss appears, we assume the LLC ways for DDIO is over-provisioned and thus will go to the Reclaim state (\circled{6}). 
Meanwhile, fewer DDIO hits and stable or even more DDIO misses indicate that core is contending LLC, so we go to the Core Demand state (\circled{7}).
Also, the High Keep state obeys the same rule (\circled{11} and \circled{12}). 

We keep in the Reclaim state if we do not observe a meaningful increase of DDIO miss count until we have reached the \textit{DDIO\_WAYS\_MIN} number of ways of LLC for DDIO, then we move to the Low Keep state (\circled{2}). 
Otherwise, we move to the I/O Demand state so that more LLC ways can be allocated for DDIO to amortize the pressure of intensive I/O traffic (\circled{5}).
At the same time, if we also observe a decrease in DDIO hit count, we will go to the Core Demand state (\circled{9}).

\subsection{LLC Re-alloc}
\label{sec:realloc}
After the state transition, \arch will take the corresponding actions, \ie, re-allocate LLC ways for DDIO or cores. 

First, \arch changes the number of LLC ways that are assigned to DDIO or tenants.
In the I/O Demand state, \arch increases the number of LLC ways for DDIO by one per iteration. 
In the Core Demand state, \arch increases the number of LLC ways for the selected tenant by one per iteration.
In Low Keep and High Keep states, \arch does not change the LLC allocation. 
In the Reclaim state, \arch reclaims one LLC way from DDIO or core per iteration, depending on the values it observes (\eg, smaller LLC miss count of the system or smaller LLC reference count of a tenant).

\arch should identify the workload that requires more or fewer LLC ways in the Core Demand and Reclaim states. The mechanism depends on the models of the end-host network we are applying. 
In the aggregation model, all the Rx/Tx buffers are allocated and managed by the virtual switch (\eg, OVS). 
This means a performance drop of the virtual switch can bottleneck the performance of the networking applications running in the attached tenants.
So, in this case, \arch increases/decreases the number of LLC ways for virtual switch's cores at first. 
In the slicing model, however, the Rx/Tx buffers of each VF are managed by the tenants themselves. 
\arch selects the tenant that needs more LLC ways from all network-related tenants by sorting their delta of LLC miss rate (in percentage) between the current and the last iteration and chooses the one with the most LLC miss rate increase. 
In this way, we are able to satisfy the LLC demand of the corresponding cores, and thus reduce the DDIO miss.

Second, \arch will shuffle the LLC ways that have been assigned to tenants, \ie, properly select the tenants that share LLC ways with DDIO.
As we discussed in \secref{sec:lc}, sharing LLC ways with DDIO can incur performance degradation of the core, even if the core is running a non-networking workload. 
Hence, it is necessary to reduce such interference. 
First of all, we should avoid any core-I/O sharing of LLC ways, if LLC ways have not been fully allocated. 
If sharing is necessary, intuitively, the tenants running PS workloads (high priority) should be isolated from LLC ways for DDIO as much as possible. 
So \arch tries the best only to overlap the LLC ways for best-effort tenants with DDIO. 
At the same time, we do not want the BE tenants to contend LLC with DDIO too much since the PS tenants' performance is correlated to DDIO's~\cite{tootoonchian2018resq,ousterhout2019shenango}. 
So before shuffling, \arch sorts all the best-effort tenants by their LLC reference count in the current iteration, and choose the one(s) with the smallest value to share LLC ways with DDIO.

\subsection{Sleep}
Since the LLC needs to be re-filled in order to be utilized, the effect of LLC allocation may take some time to show up after the \texttt{LLC Re-alloc} step. 
Also, polling the performance counters are not for free -- it suffers from the overhead of accessing the corresponding registers and data processing (see \secref{sec:overhead}).
Hence, it is necessary to select a proper polling interval for \arch. 
During this interval, \arch will simply sleep to let the OS schedule other tasks on the core. 
After each \texttt{Sleep}, if \arch is informed about changes of tenants (\eg, tenant addition/removal, change from non-networkinng to networking), it will go through the \texttt{Get Tenant Info} and \texttt{LLC Alloc} steps. 
Otherwise, it will conduct the next iteration of \texttt{Poll Prof Data}.

\begin{figure}[!t]
    \begin{subfigure}[b]{\linewidth}
        \centering
      \includegraphics[width=0.7\linewidth]{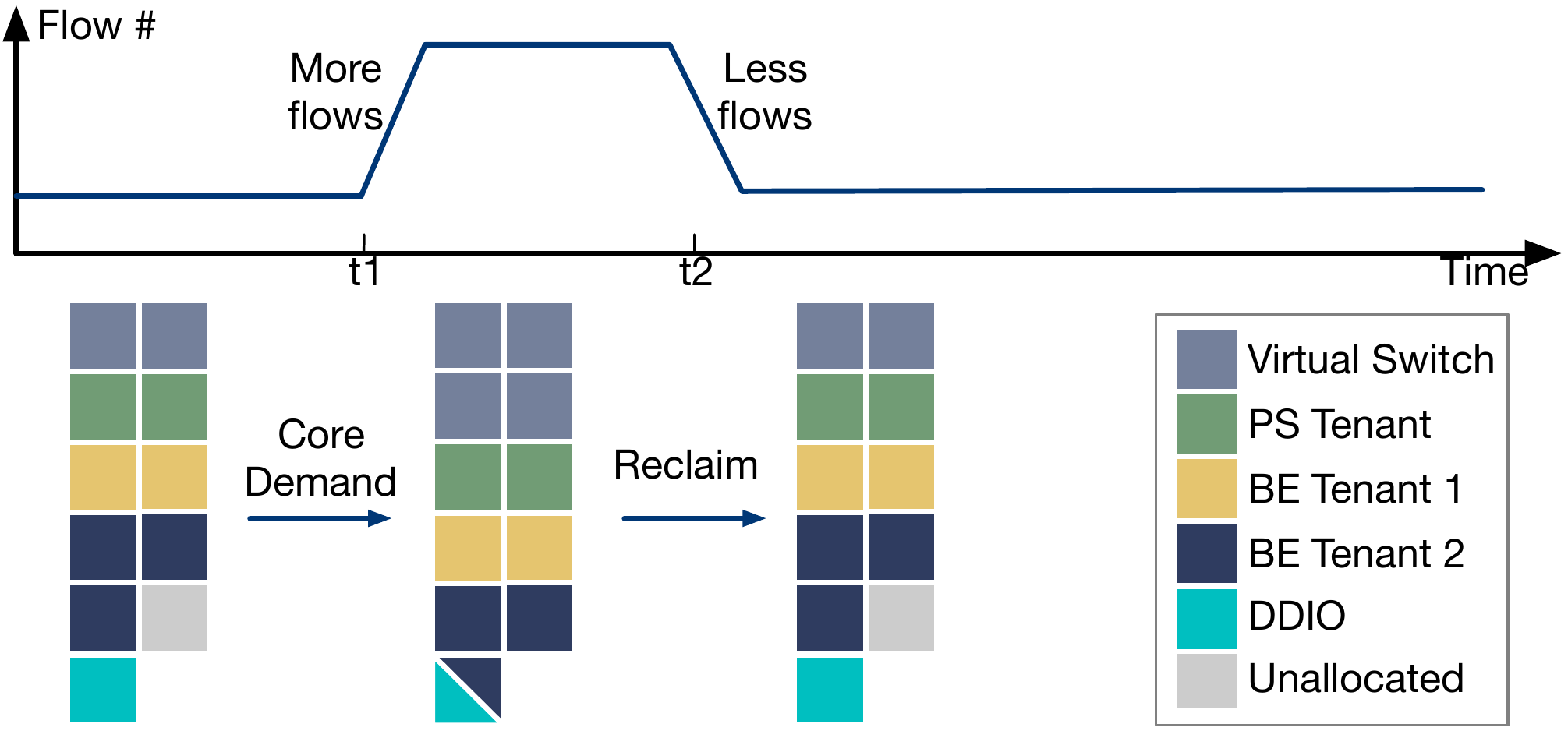}
        \caption{Example 1 with aggregation model.}
      \label{fig:e1}
    \end{subfigure}
     \hfill
    \begin{subfigure}[b]{\linewidth}
        \centering
      \includegraphics[width=0.7\linewidth]{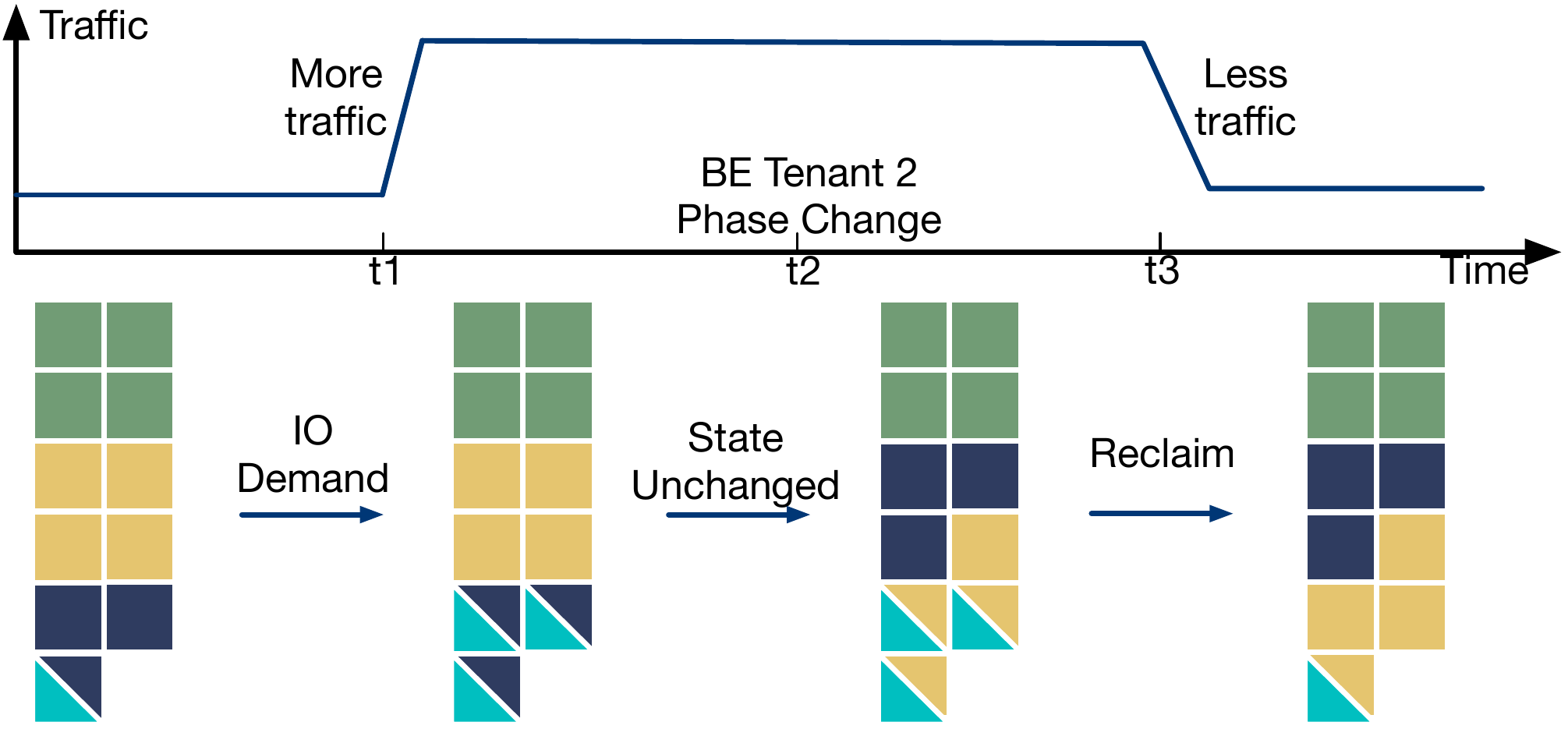}
      \caption{Example 2 with slicing model.}
      \label{fig:e2}
    \end{subfigure}
     \vspace{-4ex}
     \caption{Two examples of LLC allocation with \arch.}
      \label{fig:example}
     \vspace{-4ex}
\end{figure}

\subsection{Putting it Together: Two Examples}
We use two tangible examples to illustrate how \arch works (see \figref{fig:example}).
In the first example (\figref{fig:e1}), we use the aggregation model for the end-host network, and the throughput of network traffic is fixed. 
We have three tenants, one PS and two BE. 
Each tenant is assigned with different LLC ways. 
In the beginning, the flow count of the network traffic is small, and BE Tenant 2 shares LLC ways with DDIO. 
At the time \textit{t1}, a large number of flows appear in the traffic. 
As a result, the flow table in the virtual switch becomes larger and requires more LLC capacity than the two LLC ways that are already assigned. 
Hence, \arch detects more DDIO misses and fewer DDIO hits, and goes to the Core Demand state. 
Then, two more LLC ways are assigned to the virtual switch (one for each iteration), so that the system reaches a new balance. 
To make room for the virtual switch, we shift the LLC ways of other tenants and let BE Tenant 2 share LLC ways with DDIO. 
At the time \textit{t2}, many flows have ended, and there is no need for the virtual switch to maintain a big flow table in the LLC. 
\arch goes to the Reclaim state and reclaims two LLC ways from the virtual switch. 
Also, since now we have idle LLC ways, we remove the core-I/O sharing of LLC ways.

In the second example (\figref{fig:e2}), we assume the slicing model with the same tenants setup, and the throughput of the network traffic begins from a low intensity. 
At the time \textit{t1}, more traffic comes into PS Tenant, and the number of LLC ways for DDIO becomes insufficient, which leads to more DDIO misses. 
\arch detects this situation and transits to the I/O Demand state,  allocating more LLC ways for DDIO to achieve the balance. 
At the time \textit{t2}, the workload in BE Tenant 2 enters a new phase, which is LLC-consuming. 
\arch notices this by the delta of LLC reference count and let BE Tenant 1, which consumes less LLC, share the LLC ways with DDIO to reduce the performance interference. 
At the time \textit{t3}, the amount of incoming network traffic is decreasing, and the LLC ways for DDIO is more than enough. 
Thus, \arch can reclaim some LLC ways from DDIO.

%% file: implementation.tex
\section{Implementation}
We implement \arch as a user-space daemon, which is transparent to the application and the OS. 
Currently, we choose user-space implementation because it is more portable and flexible.
However, \arch can be implemented as a kernel module as well. 
Since the x86 instructions for MSR manipulation (\texttt{rdmsr} and \texttt{wrmsr}) are ring-0, a kernel-space implementation can potentially have lower monitoring and control overhead. 
Note that \arch can also be integrated into other CPU resource management systems~\cite{xu2018dcat,park2019copart,xiang2018dcaps}.

\niparagraph{LLC allocation.} For standard CAT functionalities (\ie, allocating LLC ways for cores), we leverage APIs from the Intel \texttt{pqos} library~\cite{pqos}. 
To better isolate different applications and demonstrate the influence of DDIO, we do not overlap LLC ways across different tenants (but it can be explored~\cite{196286,xiang2018dcaps}).
For changing and querying the LLC ways for DDIO, we write and read the DDIO-related MSRs (documented in ~\cite{sdm}) via the \texttt{msr} kernel module.

\niparagraph{Profiling and monitoring.}
Similarly, for normal profiling and monitoring (LLC miss, IPC, \etc), we use \texttt{pqos}'s APIs. 
For monitoring DDIO's hit and miss, we use the uncore performance counters~\cite{uncore-pmu}. 
It is worth noting that modern Intel CPUs apply the non-uniform cache access (NUCA) architecture~\cite{huh2005nuca,kim2002adaptive} to physically split the LLC into multiple slices. 
To reduce the monitoring overhead, for each DDIO event, we only use the performance counters in the Caching and Home Agent (CHA, the controller of each LLC slice in Intel CPUs) of one LLC slice. 
Since modern Intel CPUs apply a hashing mechanism for LLC addressing~\cite{Maurice:2015:REI:2939207.2939211,Liu:2015:LCS:2867539.2867673,Irazoqui:2015:SRE:2859854.2860454}, the data (from both the core and the DDIO) is distributed to all the LLC slices evenly. 
Hence, by only accessing one LLC slice's performance counters, we can infer the full picture of the DDIO traffic by multiplying it by the number of slices.

\niparagraph{Tenant awareness.} Since CAT assigns LLC ways to cores, we need to know the core affiliation of each tenant, or specifically, each container. 
For simplicity, we keep such affiliation records in a text file. 
When the daemon is starting, it will first read and parse the records from this file. 
Note that in the real-world cloud environment, \arch can have an interface to the container orchestrator or scheduler (\eg, Kubernetes~\cite{kubernetes}), so that it can dynamically query the affiliation information.

%% file: evaluation.tex
\section{Evaluation}
\label{sec:eval}
\subsection{Setup}
\label{sec:setup}
\niparagraph{Hardware.} We do experiments on a quad-socket Intel server with Xeon Scalable Gold 6140 CPUs~\cite{6140} (Hyper-Threading disabled). 
The CPU configuration is shown in \tabref{tab:cpu}. 
The server has 512GB DDR4 memory and two Intel XL710 40GbE NICs~\cite{nic} (both attached to socket-0). 
We connect each NIC directly to another server as the traffic generator.

\niparagraph{System software.} To reflect the multi-tenant cloud environment, we run applications in docker containers. 
For network connectivity, we have two models. 
(1) \textit{Aggregation}: we connect the physical NICs and containers via Open vSwitch~\cite{ovs2} (DPDK-based). 
(2) \textit{Slicing}: we bind a VF of the physical NIC to each container with SR-IOV technique. 
By default, we use 1024 entries as the Rx/Tx buffer size. 
For the containers that require a TCP/IP stack, we use DPDK-ANS~\cite{dpdk-ans} for high performance. 
Both the host and the containers run Ubuntu 18.04.
To not disturb the execution of tenants' application, we run \arch daemon on a dedicated core.

\begin{table}[!tb]
    \centering
    \caption{Configuration of Intel Xeon 6140 CPU.}
    \vspace{-1ex}
  \scriptsize
    \label{tab:cpu}
    \begin{tabular}{l|l}
      \hline
     Cores &  18 cores, 2.3GHz\\
      \hline
     Caches &  \tabincell{l}{8-way 32KB L1D/L1I,\\ 16-way 1MB L2,\\ 11-way 24.75MB non-inclusive shared LLC\\ (split to 18 slices)}\\
      \hline
      Memory & 6 DDR4 channels \\
      \hline
    \end{tabular}
    \vspace{-3ex}
\end{table}
\begin{table}[!tb]
  \centering
  \caption{\arch parameters.}
  \vspace{-1ex}
\scriptsize
  \label{tab:para}
  \begin{tabular}{l|l}
    \hline
   Name &  Value\\
    \hline
    THRESHOLD\_STABLE & 3\% \\
    \hline
    THRESHOLD\_MISS\_LOW & 1M \\
    \hline
    DDIO\_WAYS\_MIN/MAX &  1/6\\
    \hline
    Sleep interval & 1 second\\
    \hline
  \end{tabular}
  \vspace{-3ex}
\end{table}

\niparagraph{\arch parameters.} We use empirical parameters listed in \tabref{tab:para}. They can be tuned based on various QoS requirements and hardware configurations.

\subsection{Microbenchmark Results}

\begin{figure}[!t]
  \centering 
  \includegraphics[width=\linewidth]{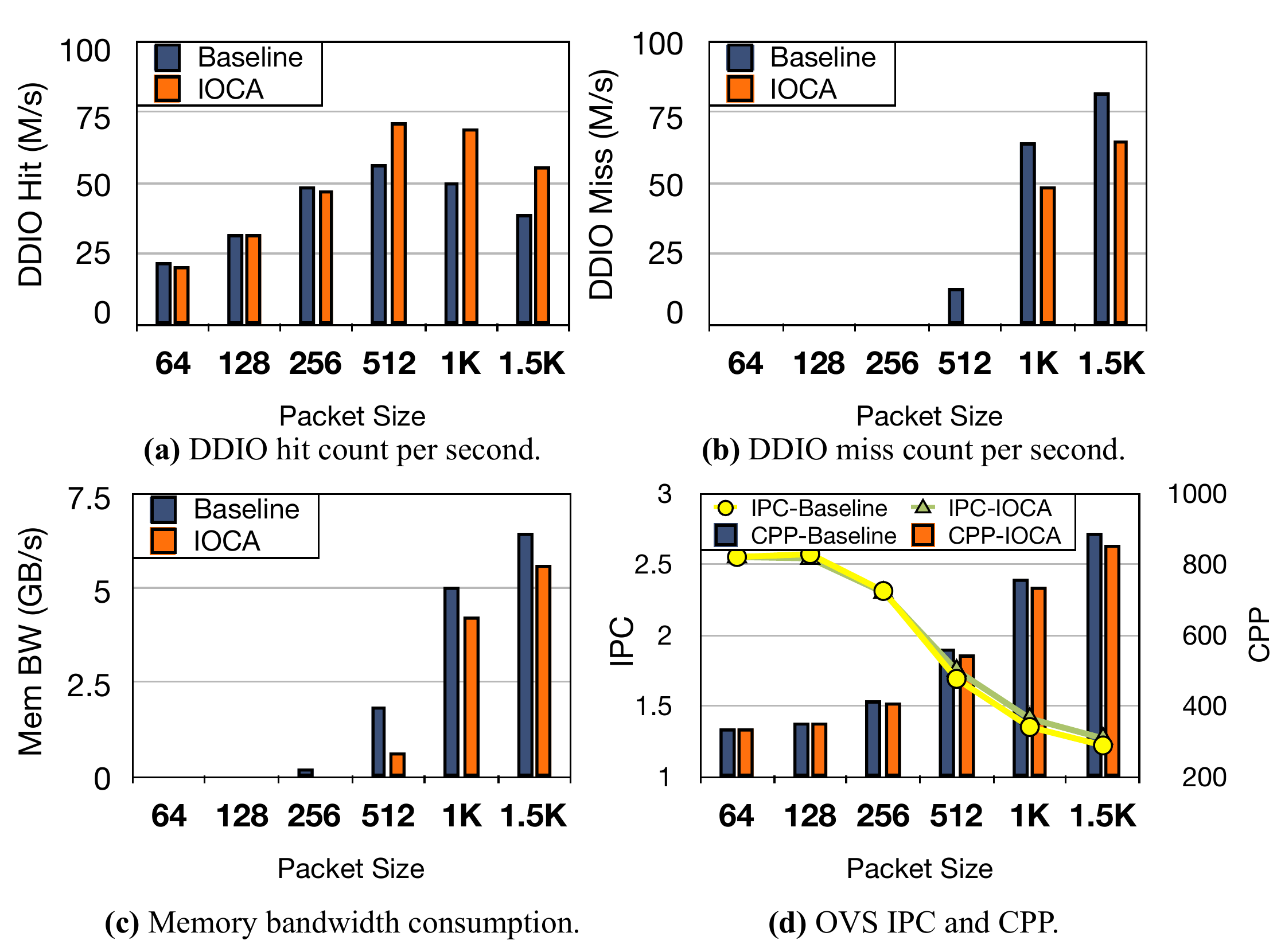}
  \vspace{-3ex}
  \caption{System performance with different packet sizes.}
  \vspace{-4ex}
  \label{fig:data5}
\end{figure}

\niparagraph{Solving the Leaky DMA problem.}
We verify whether \arch can effectively alleviate the Leaky DMA problem described in \secref{sec:ld}. 
For the aggregation model, we set up two containers running DPDK \textit{test-pmd}, each with two dedicated cores and one dedicated LLC way, both connected to OVS, which is running on two dedicated cores and two dedicated LLC ways. 
Also, the two NICs are connected to OVS. 
To reduce the overhead of other factors, we insert four rules to OVS, ``NIC0->Container0'', ``NIC1->Container1'', ``Container0->NIC0'', and ``Container1->NIC1''.
Both NICs send single-flow traffic with line rate. With such settings, the LLC miss of OVS itself is negligible and thus will not affect the performance. 
We start the experiment from 64B packet size, and over time, when the performance of OVS gets stable, we double the packet size until the MTU size (1.5KB). 

We collect the performance numbers of baseline (\ie, default DDIO configuration without \arch, but with basic CAT for cores) and \arch case and show them in \figref{fig:data5}. The most essential results are DDIO hit count (\figref{fig:data5}a) and miss count (\figref{fig:data5}b). 
When the packet size is small, the default two LLC ways for DDIO are enough to contain the on-the-fly inbound packets, which means there is no need for \arch to change the state and allocate more LLC ways for DDIO; however, as packet size increases over time, on-the-fly packets put much more pressure on the LLC, and thus the default two LLC ways become deficient, which is reflected on the increase of DDIO miss count. 
At this time, \arch detects the unstable status and transits its state to I/O Demand, and thus allocates more LLC ways for DDIO (one by each time). 
As a result, the DDIO hit count with \arch is higher than the baseline, and the DDIO miss count is lower. 
This leads to better memory throughput performance in \figref{fig:data5}c. We can find that with \arch, the memory bandwidth consumption can be reduced by at most 15.6\%. 
Note that since the limited capacity of LLC, \arch is not able to eliminate the memory traffic. 
It is desirable to combine \arch and a slightly smaller Rx buffer (\eg, 512 in \figref{fig:data1}) to achieve even better memory traffic reduction with modest throughput loss. We also plot OVS performance in \figref{fig:data5}d. 
Instruction per cycle (IPC) and cycle per packet (CPP) are measured. 
Again, with large packet sizes, \arch is able to improve OVS's IPC by $\sim5\%$ and also reduce CPP. 
This improvement requires no software/hardware modification and can be even more significant with much higher bandwidth NIC in the future.

\begin{figure}[!t]
  \centering 
  \includegraphics[width=\linewidth]{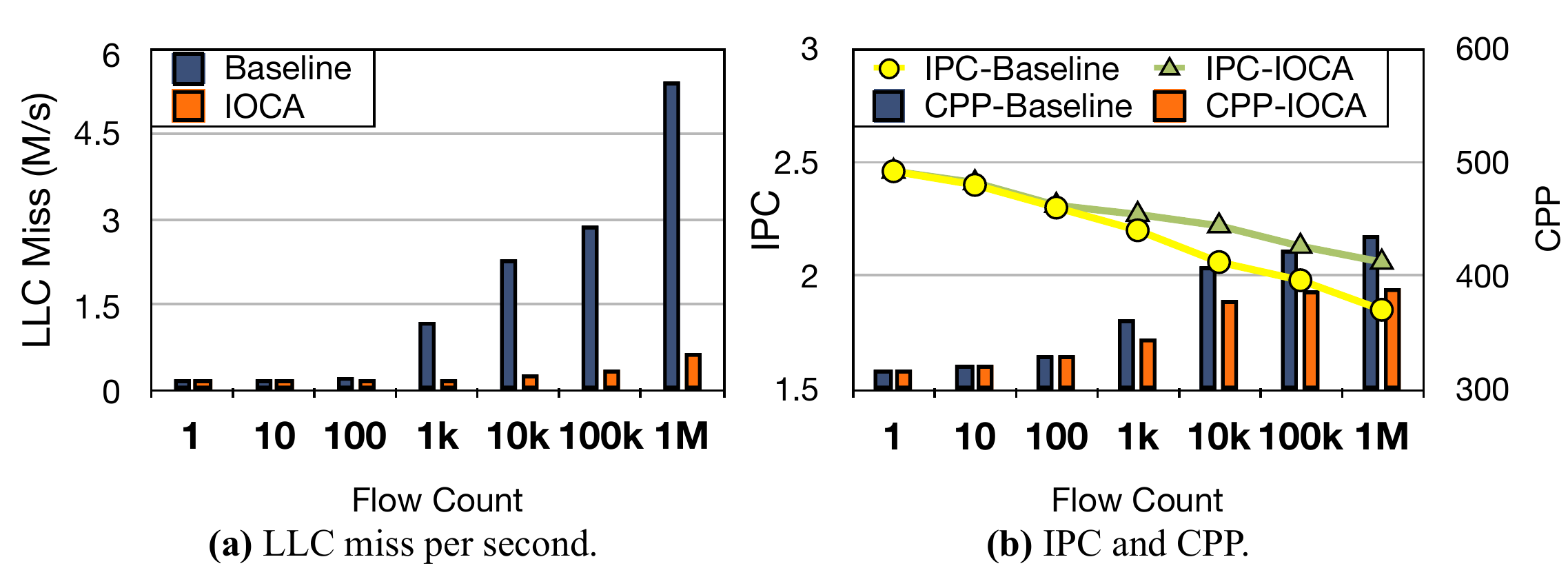}
  \vspace{-4ex}
  \caption{OVS performance with different flow counts.}
  \vspace{-3ex}
  \label{fig:data7}
\end{figure}

At the same time, \arch can still identify the core's demand for LLC capacity in a networking application. We demonstrate this with a second experiment with similar settings. 
The difference is, we fix the packet size at 64B (so that cores will be the dominant source of LLC miss). 
We start the line-rate traffic from single flow, and gradually increase the number of flows in the traffic, and report the performance in \figref{fig:data7}. 
To maintain the growing number of flows in its internal flow table, OVS requires an increasing amount of memory space.
Hence, if we keep the static initial LLC allocation for OVS, it will suffer from higher LLC miss count after more than 1k flows, and thus lower IPC. 
On the other hand, \arch, by detecting the drop of IPC and the increase of LLC miss rate, is able to identify the demand for LLC of OVS's cores and allocate more LLC ways for these cores. 
As a result, OVS maintains a low LLC miss count and gains at most 11.4\% higher IPC than the baseline. 
Note that with more flows, the IPC and CPP inevitably get worse since OVS's design~\cite{ovs2} leads to more (slower) wildcarding lookups instead of pure (faster) exact match lookups.

\begin{figure}[t]
  \centering
  \begin{subfigure}[b]{0.49\linewidth}
    \centering
    \includegraphics[width=\linewidth]{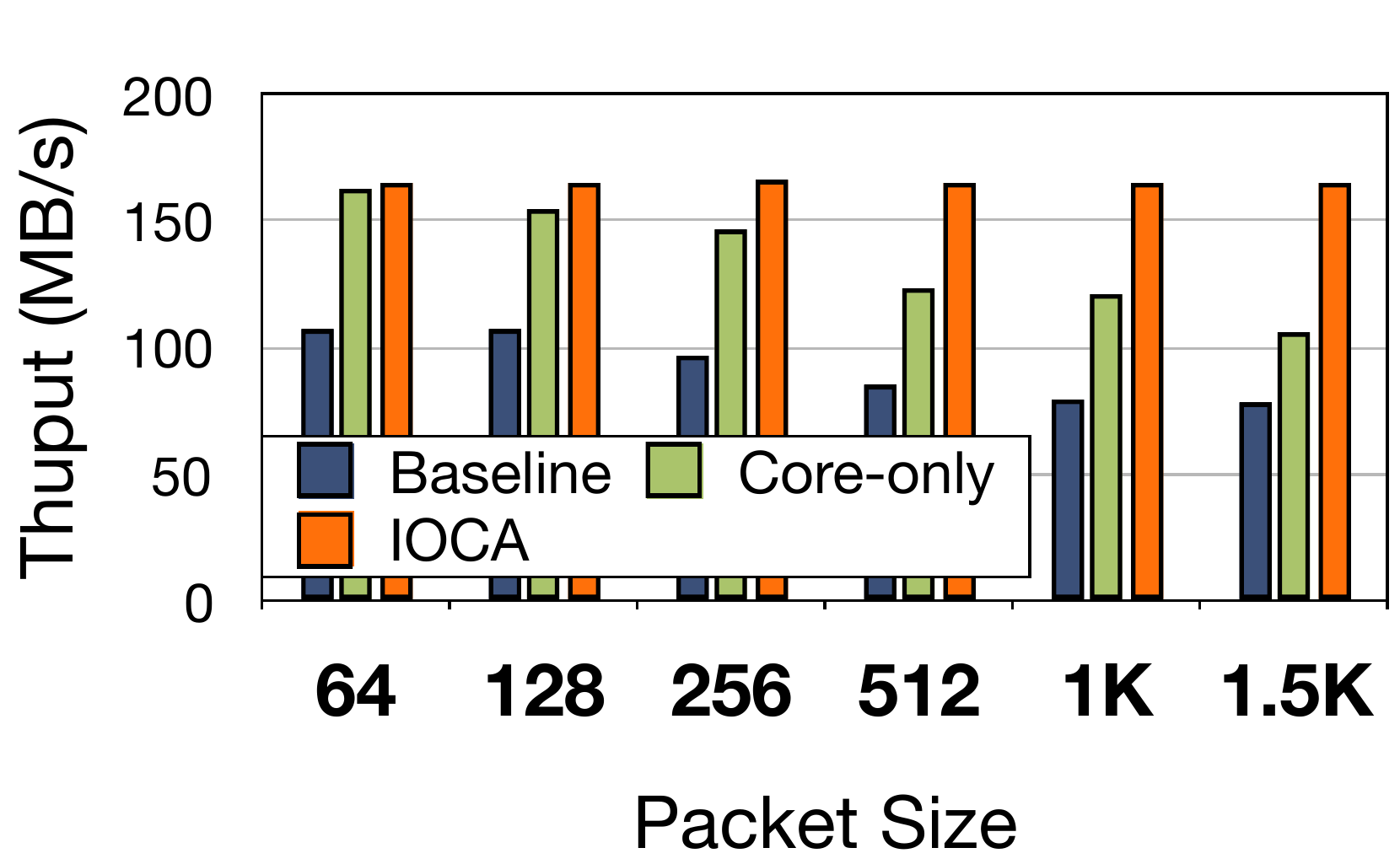}
    \vspace{-2ex}
      \caption{Throughput between t1 and t2.}
    \label{fig:data6-1}
  \end{subfigure}
   \hfill
     \begin{subfigure}[b]{0.49\linewidth}
    \centering
    \includegraphics[width=\linewidth]{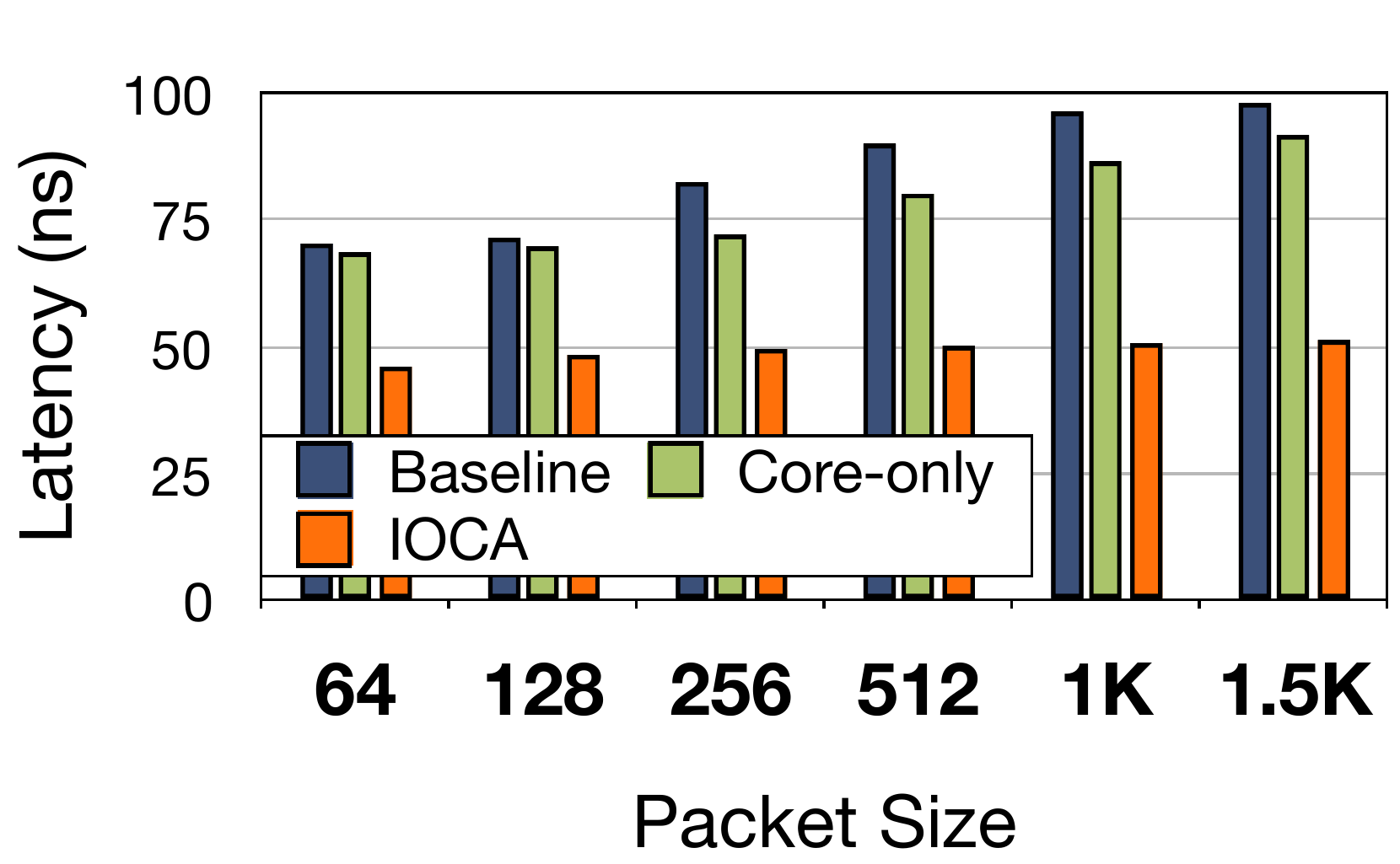}
    \vspace{-2ex}
      \caption{Avg latency between t1 and t2.}
    \label{fig:data6-2}
  \end{subfigure}
  \\
\begin{subfigure}[b]{0.49\linewidth}
    \centering
    \includegraphics[width=\linewidth]{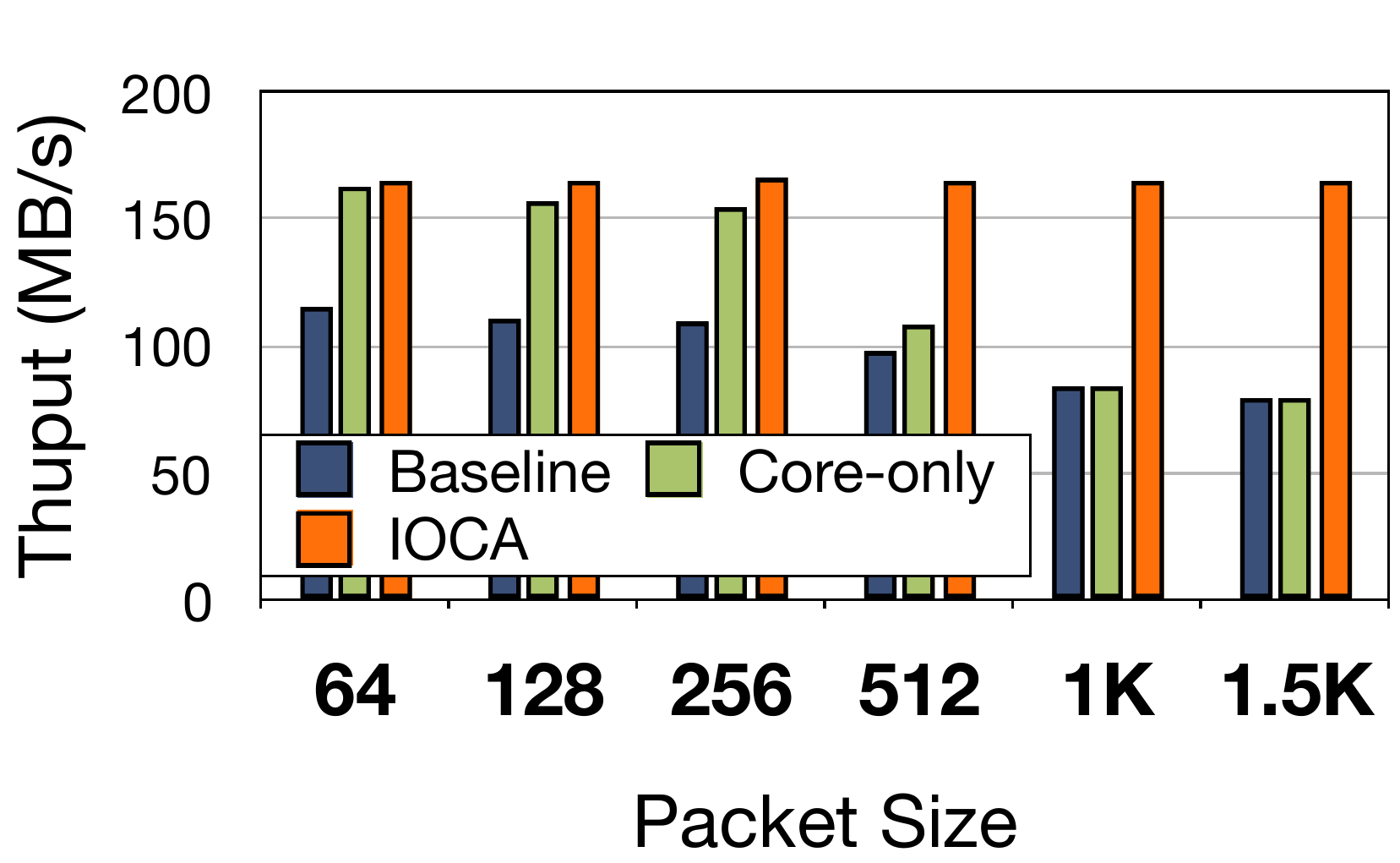}
    \vspace{-2ex}
    \caption{Throughput after t2.}
    \label{fig:data6-3}
  \end{subfigure}
   \hfill
  \begin{subfigure}[b]{0.49\linewidth}
    \centering
    \includegraphics[width=\linewidth]{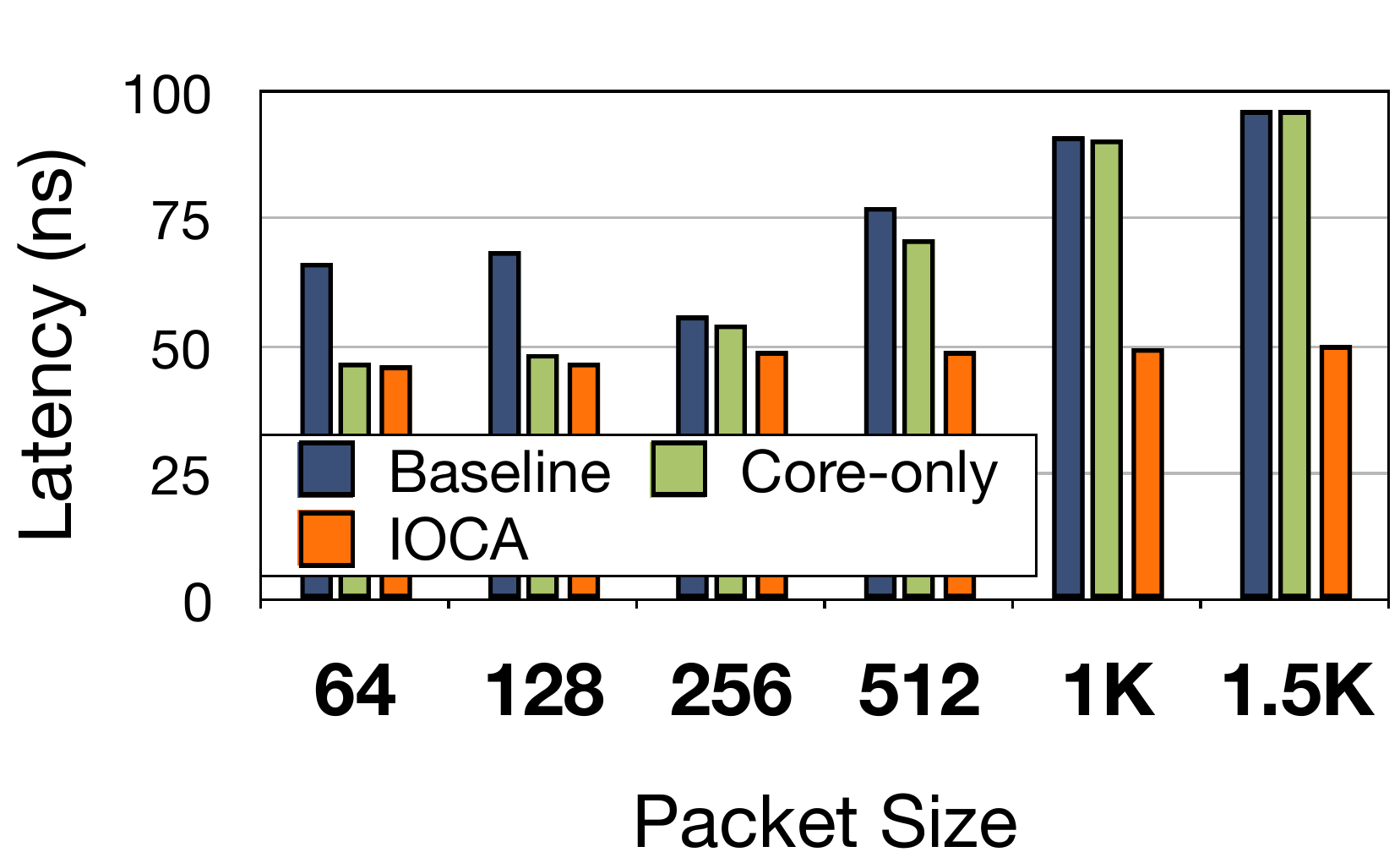}
    \vspace{-2ex}
    \caption{Avg latency after t2.}
    \label{fig:data6-4}
  \end{subfigure}
   \vspace{-2ex}
   \caption{The performance of X-Mem in container 4.}
    \label{fig:data6}
   \vspace{-2ex}
\end{figure}

\begin{figure}[!t]
  \centering 
  \includegraphics[width=0.7\linewidth]{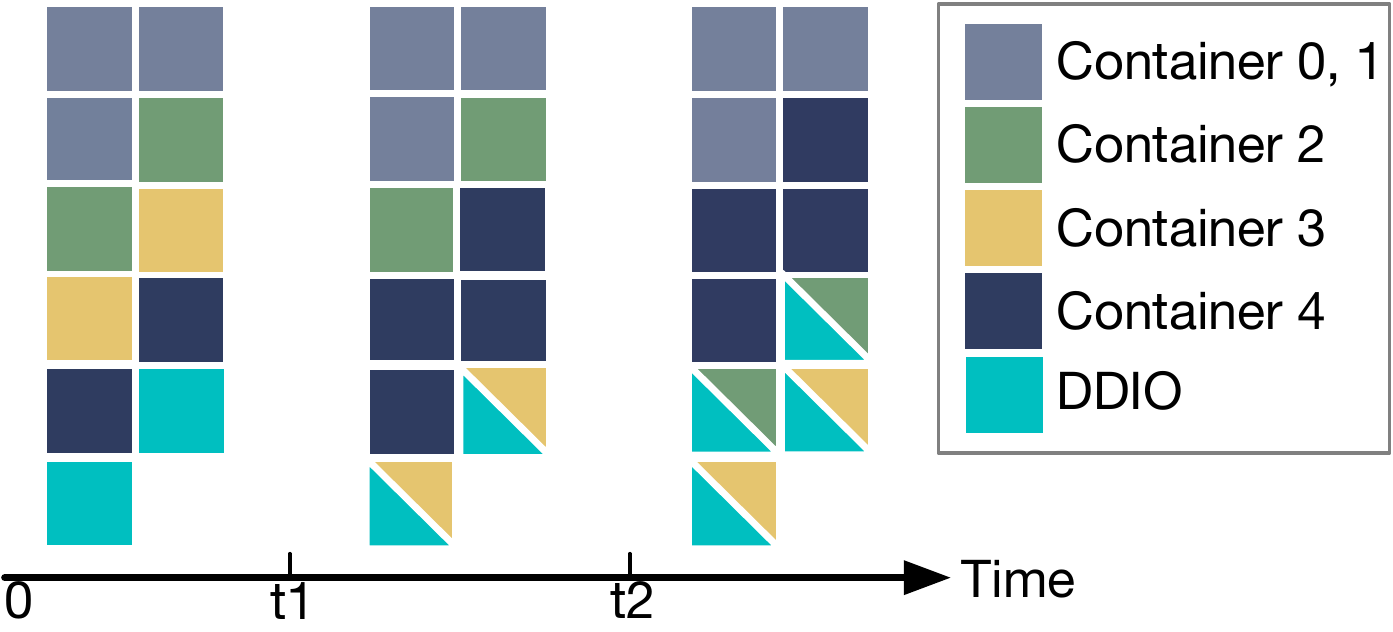}
  \vspace{-2ex}
  \caption{\arch LLC allocation for the five containers.}
  \vspace{-4ex}
  \label{fig:data6-5}
\end{figure}

\niparagraph{Solving the Latent Contender problem.\footnote{In this experiment, we temporarily disable \arch's functionality of changing LLC ways for DDIO.}}
With the slicing model, we demonstrate whether \arch can efficiently choose the LLC sharing policy between core and I/O. 
We have one VF for each NIC and bind them to two containers (0 and 1, marked as PS) running DPDK \textit{test-pmd}. 
Each container runs on one dedicated core, and they share three dedicated LLC ways (no DDIO overlap). 
On each NIC, we generate single-flow line-rate traffic with different packet sizes. 
Packets in \textit{test-pmd} will be bounced back to the NIC as outbound traffic. 
Besides, we have three identical containers (2 and 3 as BE, 4 as PS), each with one dedicated core and two dedicated LLC ways (no DDIO overlap) running X-Mem (random read pattern). 
In the beginning, all X-Mem containers have a working set of 2MB. 
At time \textit{t1}, we increase the working set size of container 4 to 10MB (L2 cache size + 4 LLC ways size). 
Furthermore, at time \textit{t2}, when the system has been stable, we manually increase LLC ways count for DDIO from two to four, and wait for the system to become stable again. 
Besides the baseline and \arch, we also test a case named Core-only, which means we only adjust the LLC allocation without I/O awareness\footnote{We do this by disabling I/O Demand state and LLC shuffling.}. This is to emulate the behavior of other state-of-the-art LLC management mechanisms for comparison.
We report the performance of X-Mem in container 4 in \figref{fig:data6}. 
During this procedure, we find after \textit{t1}, when the working set size of container 4 increases dramatically, \arch starts allocating more LLC ways for container 4, which are shared with DDIO. 
To avoid contention between core and I/O, \arch shuffles the assigned LLC ways for container 4 and select container 3 with BE workload to share LLC ways with DDIO (see \figref{fig:data6-5}). 
As seen from \figref{fig:data6-1}, larger packet sizes will put higher pressure on the LLC ways for DDIO, interfering with the core more seriously, and thus drag down the X-Mem's throughput. 
Core-only, by simply allocating two more ``idle'' (but actually shared with DDIO) LLC ways for X-Mem, performs well with small packet sizes, but fails to maintain this trend with larger packet sizes since the core-I/O contention is mitigated but not eliminated. 
On the other hand, \arch is able to maintain constantly high throughput with all packet sizes (53.6\%-111.5\% compared to baseline and 1.4\%-56.0\% compared to Core-only) since it not only allows X-Mem to use more LLC space but also avoids core-I/O contention. 
In terms of latency (\figref{fig:data6-2}), we may find Core-only does not help much with any packet size since the randomly accessed data can be in the two X-Mem-dedicated LLC ways, the two core-I/O shared LLC ways, or memory. 
On average, the latency will not be much better than doing nothing (\ie, baseline). 
However, \arch still maintains low latency ($34.5\%\sim52.2\%$ compared to baseline and $32.9\%\sim44.2\%$ compared to Core-only) regardless of the packet size since it achieves 100\% LLC isolation for container 4.   

After \textit{t2}, since DDIO is sharing (two) LLC ways with container 4 again, \arch detects the unstableness of the system by the increasing LLC miss count of container 4's core and reshuffles the LLC ways allocation so that the 100\% LLC isolation is still maintained. 
Core-only, sharing all its four LLC ways with DDIO, suffers from more severe performance interference compared to it during \textit{t1} and \textit{t2}, this is especially significant when packet size is large. Both throughput (\figref{fig:data6-3}) and latency (\figref{fig:data6-4}) are very close to the baseline. 
It is worth noting that with small packet sizes, Core-only performs better in \figref{fig:data6-4} than in \figref{fig:data6-2}, which is because the LLC ways for DDIO in \figref{fig:data6-4} are more than enough, and the inbound packet can be distributed to a larger space, which amortizes the contention.

\begin{figure*}[!t]
  \centering 
  \includegraphics[width=0.85\linewidth]{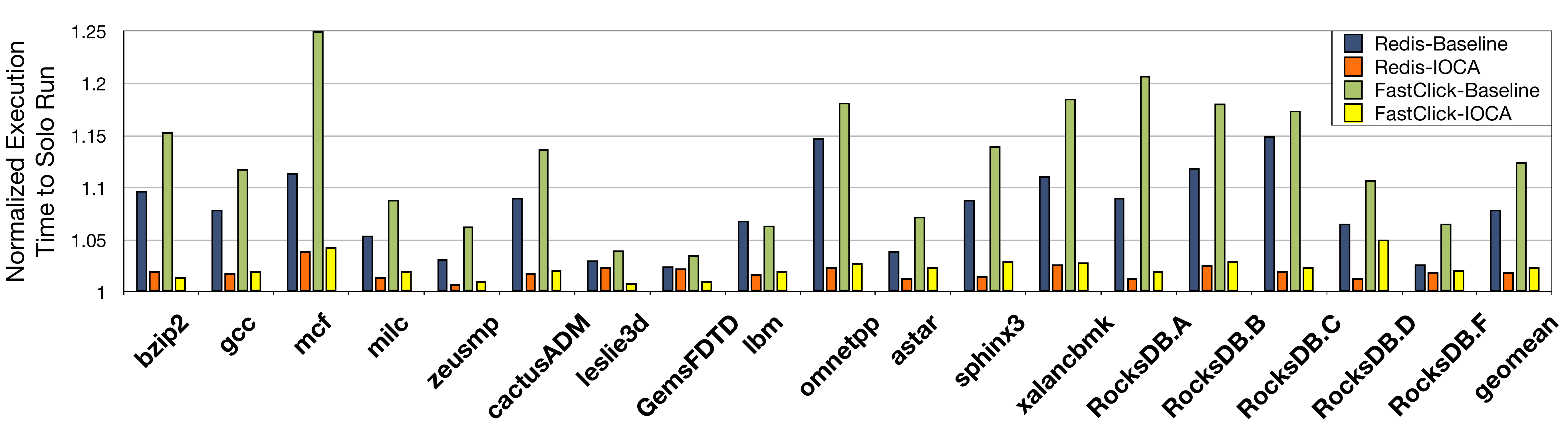}
  \vspace{-3ex}
  \caption{Normalized execution time to solo run of SPEC2006 and RocksDB with two networking applications.}
  \vspace{-3ex}
  \label{fig:data8}
\end{figure*}

\subsection{Application Results}
We evaluate \arch's impact on the performance of multiple real-world applications. 
Specifically, we have two sets of applications as the non-networking cloud workloads. 
The first one is the SPEC2006 benchmark suite~\cite{spec}. 
We run selected memory-sensitive benchmarks~\cite{jaleel2010memory}, all with the \texttt{ref} input.
The second one is RocksDB~\cite{rocksdb}, a storage-centric persistent key-value store. 
We use YCSB~\cite{Cooper:2010:BCS:1807128.1807152} with 0.99 \texttt{Zipfian} distribution to test the performance of RocksDB. 
To avoid any storage I/O operations, we only load 10K records (1KB per record) so that all records are in RocksDB's memtable. 

On the other hand, we choose in-memory key-value store (KVS) and network function virtualization (NFV) service chain as two representative networking workloads since they both involve tremendous network traffic and are cache-sensitive, which are the targeted usages for DDIO and \arch.

\niparagraph{In-memory KVS.} We use Redis~\cite{redis}, a popular in-memory KVS, to conduct the experiment. 
We run two Redis containers, each with two dedicated cores, and connect them to the OVS, which is running on another two dedicated cores. 
The OVS and two Redis containers share three LLC ways (no DDIO overlap). 
Besides, we have one PS container, which is running either a SPEC2006 benchmark or RocksDB on one dedicated core and two LLC ways. 
We also have two BE containers, each with two LLC way and one dedicated core, running X-Mem random-read, but with different working sets (one 1MB, one 10MB). 
Initially, the LLC ways allocation of the three non-networking containers are randomly shuffled, and DDIO is not taken into consideration. 
The two NICs are connected to the OVS, and we run YCSB benchmarks from the traffic generator machines (each using eight threads). 
We pre-load 1M records with 1KB size and run different operations for testing.  

\niparagraph{NFV service chain.} We run a FastClick~\cite{10.5555/2772722.2772727}-based stateful service chain with and without \arch to show the benefit of \arch. 
This service chain consists of three network functions (NFs): a firewall based on Click element \texttt{classifier}, a flow stats based on Click element \texttt{AggregateIPFlows}, and a network address/port translator (NAPT). 
Each of the two NICs is virtualized to two VFs with different VLAN tags. 
We have four identical containers, each bind to one VF (\ie, each container processing one VLAN's traffic), running the service chain on four separate and dedicated cores. Such four containers share three LLC ways (no DDIO overlap). 
The non-networking workloads are the same as the KVS experiment.
We generate traffic (all 1.5KB packet) of four VLANs from the two traffic generator machines with equal bandwidth, \ie, 20Gbps per VLAN. 

To isolate the performance impact caused by DDIO and highlight the two problems this work is committed to tackling, we temporarily disable \arch's functionality of assigning more/less LLC ways for tenants (but the ways for different tenants will still be shuffled). 
We first run each application solely (\ie, \textit{solo run} in the following figures) to get purely isolated performance. 
We then co-run the applications in the aforementioned scenarios with and without \arch (\ie, \textit{baseline} and \arch in following figures).
We run each case for multiple times and collect the maximum performance degradation of each application. 

\begin{figure}[!t]
  \centering 
  \includegraphics[width=0.7\linewidth]{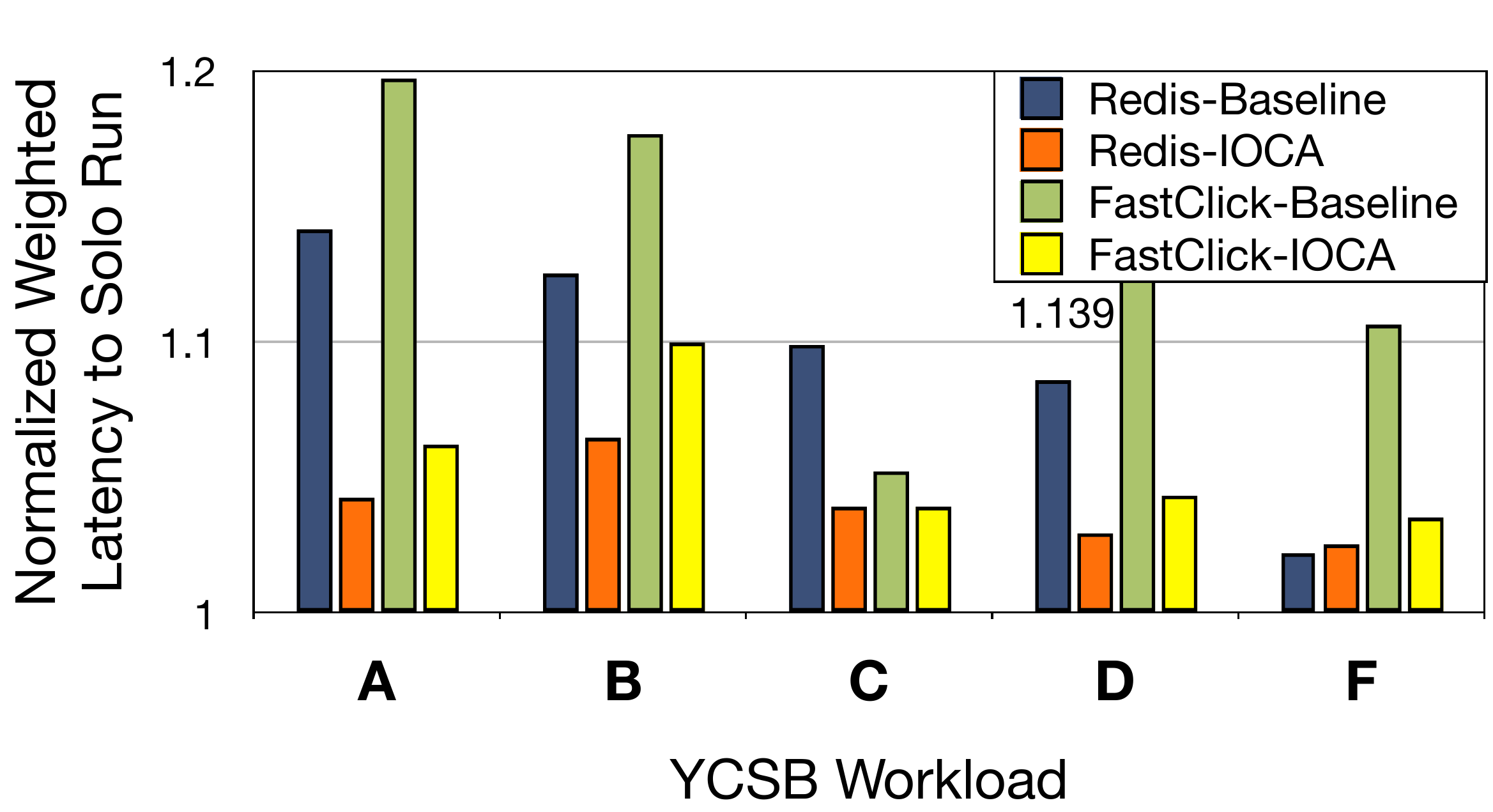}
  \vspace{-2ex}
  \caption{Normalized weighted average latency to solo run of RocksDB with two networking applications.}
  \vspace{-3ex}
  \label{fig:data9}
\end{figure}

We first report the execution time of each non-networking application normalized to solo run in \figref{fig:data8}. 
For SPEC2006, different benchmarks have different working set size and sensitivity to the cache size~\cite{jaleel2010memory}. But in general, without DDIO awareness and \arch, we can observe a $2.5\%\sim14.8\%$ performance degradation when co-running with Redis and $3.5\%\sim24.9\%$ when co-running with FastClick. 
Without DDIO awareness, a non-networking application is likely to be affected by the networking application which seems to have complete isolation against it. 
However, \arch can effectively alleviate such degradation to $0.7\%\sim3.8\%$ and $0.8\%\sim5.0\%$, respectively. 
The reasons why it is not able to perfectly match the performance of solo run are: (1) the cache needs to be warmed up to be effective when \arch shuffles the LLC ways for tenants, (2) partial LLC way overlap with DDIO may be inevitable when \arch assigns more LLC ways for DDIO (\eg, High Keep state), and (3) the memory bandwidth consumed by the networking applications may also affect the performance of non-networking applications~\cite{tootoonchian2018resq,park2019copart}. 
Applying Intel Memory Bandwidth Allocation (MBA) can solve this problem, which is out of the scope of this paper. 
Similarly, different YCSB workloads for RocksDB have different cache locality requirements and thus are affected by networking applications to various extents (\ie, $2.6\%\sim14.9\%$ and $6.5\%\sim20.6\%$, respectively). 
Again, \arch is able to shuffle the LLC ways of the non-networking application so that it is isolated from DDIO as much as possible, which leads to only $1.2\%\sim2.6\%$ and $2.0\%\sim4.9\%$ performance (throughput) degradation, respectively. 
Also note that, with more intensive network traffic (\ie, line-rate for both inbound and outbound traffic), FastClick generally exerts more impact on the performance of non-networking applications than on Redis. 
We expect Redis to be impacted more severely when running more Redis instances on a single server.

We also report the latency performance of RocksDB in \figref{fig:data9}. 
Since there can be more than one type of operation in a single YCSB benchmark, we normalize the latency of each operation and calculate the weighted average value (\ie, normalized weighted latency).
Since the key-value data of RocksDB can be evicted from the LLC to the main memory by the inbound DDIO data, the average latency performance can be much worse than solo run (\ie, as high as for 14.1\% Redis and 19.7\% for FastClick). 
On the other hand, \arch can help mitigate such unexpected eviction by shuffling the LLC ways for the non-networking application, resulting in at most 6.4\% and 9.9\% longer latency, respectively.

\begin{figure}[t]
  \centering
  \begin{subfigure}[b]{0.45\linewidth}
    \centering
    \includegraphics[width=\linewidth]{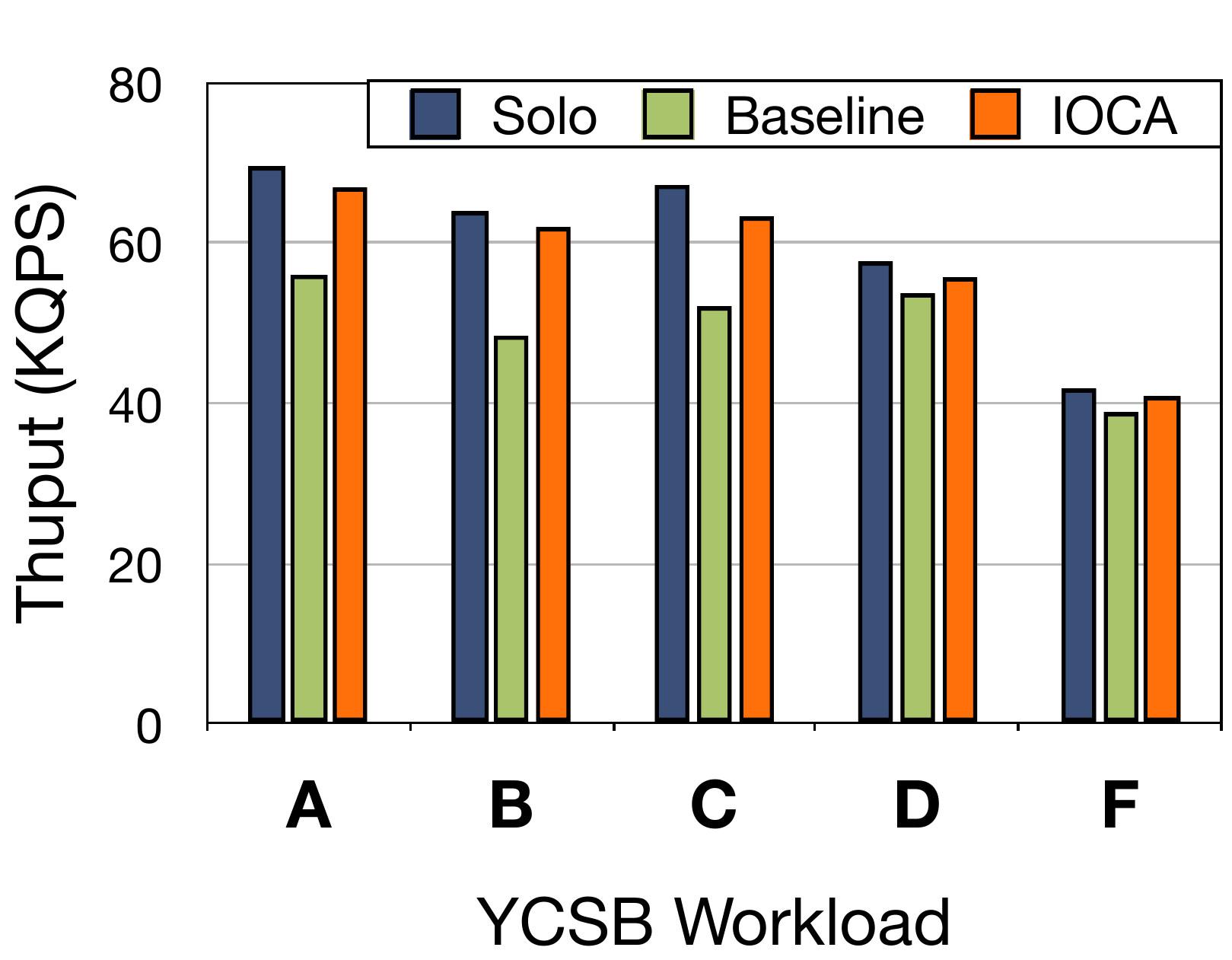}
      \caption{Throughput.}
    \label{fig:data10-1}
  \end{subfigure}
   \hfill
  \begin{subfigure}[b]{0.48\linewidth}
    \centering
    \includegraphics[width=\linewidth]{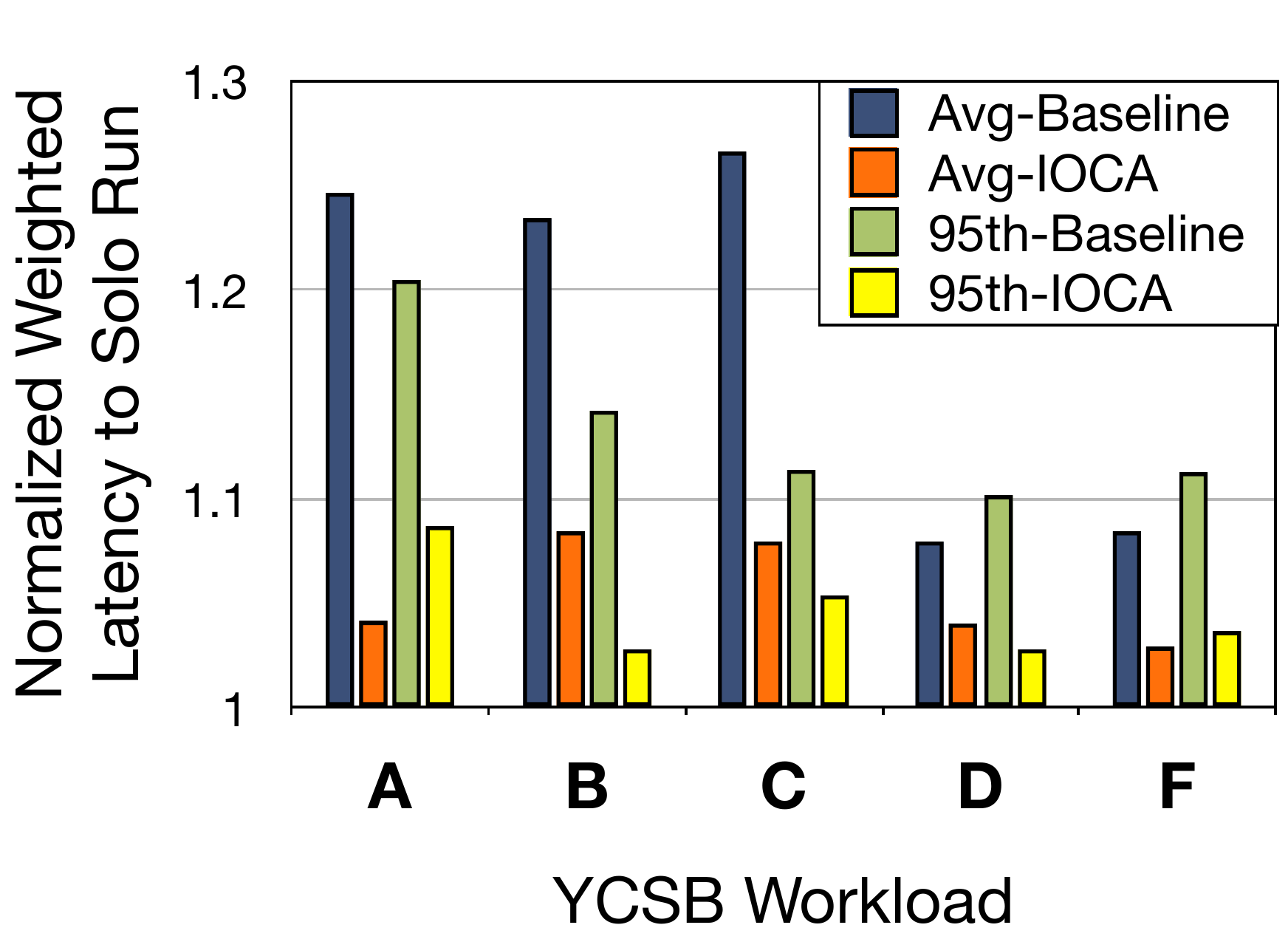}
    \caption{Latency.}
    \label{fig:data10-2}
  \end{subfigure}
   \vspace{-1ex}
   \caption{Redis performance on different YCSB workloads.}
    \label{fig:data10}
   \vspace{-4ex}
\end{figure}

We then discuss the performance of networking applications. 
\figref{fig:data10} depicts the YCSB results of Redis. 
In the baseline, since DDIO is not considered, if an application that heavily consumes cache resource (\eg, X-Mem with 10MB working set, \texttt{mcf}, \texttt{omnetpp}, and \texttt{xalancbmk} in SPEC2006, and RocksDB) happens to be sharing LLC ways with DDIO, not only such non-networking application itself but also the networking applications will be adversely impacted. 
Specifically, we can see $7.1\%\sim24.5\%$ throughput degradation, $7.9\%\sim26.5\%$ longer average latency, and $10.1\%\sim20.4\%$ longer tail latency among different YCSB workloads,  especially with workloads that involve dense read operations (\ie, A, B, and C). 
\arch mitigates such degradation by (1) allocate more LLC ways for DDIO to inject inbound packets into the LLC, and (2) shuffling LLC ways to minimize, if not eliminate, the overlap between DDIO and cache-hungry applications. 
These two methods seem a little contradictory since more LLC ways for DDIO means more chance to overlap with other applications. 
But actually, with more LLC ways for DDIO, inbound packets can be distributed evenly among LLC ways, amortizing pressure on each single LLC way. 
Even if a few ways are overlapped, the overall benefit still outperforms the adverse impact. 
As a result, \arch minimizes the performance degradation to $2.8\%\sim5.6\%$, $2.9\%\sim8.9\%$, and $2.8\%\sim8.7\%$. 

Regarding FastClick, since we are using traffic of large packet, the CPU core is not the bottleneck of packet processing, we do not observe meaningful throughput drop of the service chain. 
Also, due to the limitation of the software packet generator~\cite{pktgen-dpdk} we are using, we are not able to report the average and tail latency. 
But we do see a lower maximum round-trip latency and fewer time variances (\ie, a large difference between the round-trip latency of two consecutive packets) with \arch, compared to baseline. 
This shows, allowing more packets to be fetched and processed from the LLC, \arch makes the performance of the FastClick service chain more stable.

\begin{figure}[!t]
  \centering 
  \includegraphics[width=0.6\linewidth]{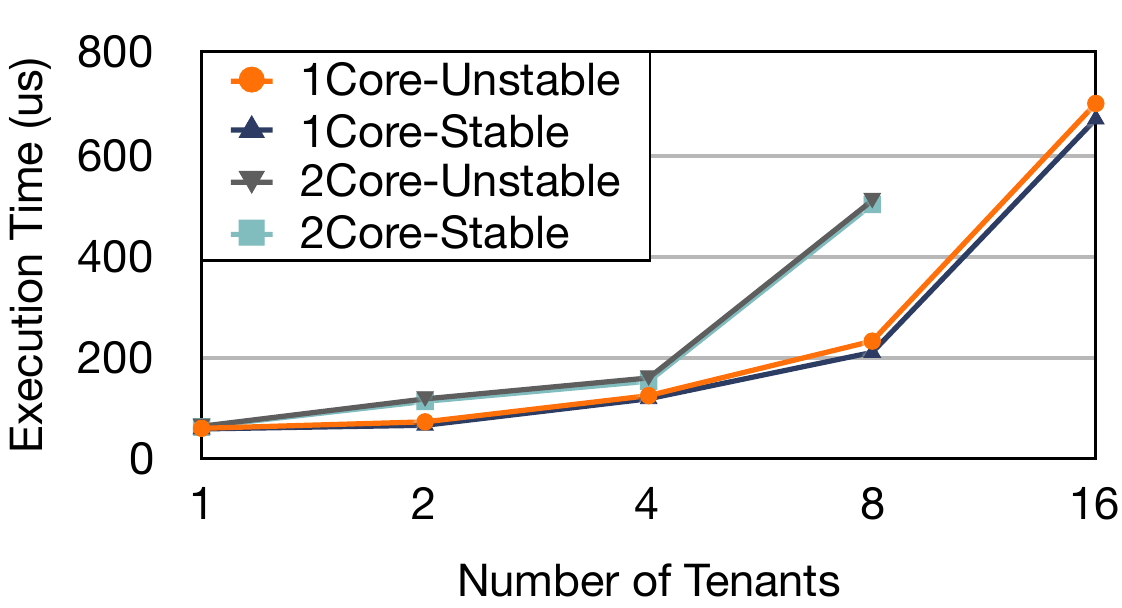}
  \vspace{-2ex}
  \caption{\arch execution time with different tenant counts.}
  \vspace{-4ex}
  \label{fig:data4}
\end{figure}

\subsection{IOCA Overhead}
\label{sec:overhead}
We finally investigate the overhead of the \arch user-space daemon. 
Specifically, we set up different numbers of tenants and measure \arch's execution time of each interval (excluding the initialization time and sleep time). 
We measure two cases: (1) each tenant has one dedicated core, and (2) each tenant has two dedicated cores. 
We run the \arch daemon on the its dedicated core for 1000 seconds and report the average values of execution time. 
We classify the results into two categories, \ie, Stable (only the \texttt{Poll Prof Data} time) and Unstable (\texttt{Poll Prof Data + State Transition + LLC Re-alloc} time), and depict them in \figref{fig:data4}\footnote{Since our CPU only has 18 cores, for the two cores per tenant case, we have at most eight tenants.}. 

First, we find that most of \arch's execution time is spent on the \texttt{Poll Prof Data} step, while conducting \texttt{State Transition} and \texttt{LLC Re-alloc} is relatively cheap. 
This is because, in \texttt{Poll Prof Data} step, the \arch daemon needs to read and write CPU hardware performance counters, each with costly context-switch (recall that the instruction that reads the counters is ring-0).
In contrast, the \texttt{State Transition} step is mainly branches and numerical comparison, and \texttt{LLC Re-alloc} typically only involves a couple of (\ie, fewer than five) CPU register writes. 

Second, \arch's execution time increases roughly sub-linearly with the number of cores it is monitoring. 
Since IPC and LLC reference/miss are per-core metrics, monitoring more cores means more counter read operations, which is the dominant part of the execution time. 
At the same time, with the same number of cores, fewer tenants correspond to a shorter time, since some of the counter read operations coalesce, and the overhead of context-switch is alleviated. 
Even with a large number of cores, \arch's execution time does not exceed 800us, which shows the lightweightness and efficiency of \arch. 
That is, given the one-second interval in this paper, \arch, if co-running on a tenant's core, may only add negligible overhead (\ie, at most 0.08\%) to the system.

%% file: discussion.tex
\section{Discussion}
\label{sec:disc}
\niparagraph{Storage.} 
As we mentioned, DDIO is a technology for all PCIe devices. 
In this paper, we mainly focus on the performance interference of LLC caused by NIC. 
However, NIC is not the only factor that can affect the LLC.
Storage, as another essential system building block, can also lead to interference.
Such interference can be especially significant 
as high-speed NVMe drives (bandwidth as high as 2.5GB/s~\cite{p4800}) being used with user-space storage stacks (\eg, SPDK~\cite{yang2017spdk}). 
We believe that \arch can be used for storage devices as well by future extension.

\niparagraph{DDIO for the remote socket.} 
Currently, DDIO only supports the local socket. 
That is, inbound packets are only injected into the socket that the corresponding I/O device is attached to, even if the application is running on a remote socket~\cite{remote-ddio}.
Hence, we are constrained to only leverage and monitor the LLC on the local socket.
One solution to overcome this constraint is to apply the multi-socket NIC technology~\cite{socketdirect,10.1145/3373376.3378509} so that the inbound packets from the same NIC can be directly dispatched to their corresponding socket. 
We also expect that DDIO can be extended to support remote sockets through socket interconnect (\ie, Intel UPI) in future CPUs.

\niparagraph{Future DDIO consideration.}
The current DDIO implementation in Intel CPUs does not distinguish among devices and applications. 
That is, inbound traffic (both \texttt{write update} and \texttt{write allocat}) from various PCIe devices is treated the same.
This, in turn, may cause performance interference between applications that use DDIO simultaneously. 
For example, a BE batch application (\eg, Hadoop) with heavy inbound traffic may evict the data of other PS applications (\eg, Redis, NFs) from DDIO's LLC ways, which leads to performance degradation of those PS applications.
However, the batch applications, whose performance is insensitive to the memory access latency, cannot get significant benefit from fetching data from the LLC instead of the memory. 
We expect that DDIO in future Intel CPUs can be device-aware. 
\Ie, it can assign different LLC ways to different PCIe devices, or even different queues in a single device, just like what CAT does on CPU cores.
And \arch can further evolve to leverage such awareness.

%% file: related.tex
\section{Related Work} 
\label{sec:related}
\subsection{Cache Partitioning and Isolation}
Both industry and academia have been focusing on cache management for a long time. 
On the hardware side, many mechanisms~\cite{qureshi2006utility,Albonesi:1999:SCW:320080.320119,Qureshi:2007:AIP:1250662.1250709,Balasubramonian:2000:MHR:360128.360153,Varadarajan:2006:MCC:1194816.1194856,Sanchez:2011:VSE:2000064.2000073,10.1145/2897937.2898036,7056026,10.1145/2872362.2872382} have been proposed for cache partitioning with different granularities.
However, most of them require significant hardware/OS changes and thus have only been evaluated in the simulation environment. 
On the software side, page coloring~\cite{Bugnion:1996:CPC:237090.237195,Zhang:2009:TPP:1519065.1519076,ye2014coloris} is one of the representative techniques. 
Page coloring requires no hardware changes, but its high overhead and low flexibility~\cite{el2018kpart,xu2018dcat} prevent it from being widely adopted.  

Intel RDT~\cite{herdrich2016cache,rdt} first provides hardware support for flexible cache resource management in practice. 
It provides software with interfaces for cache (and memory) monitoring and partitioning in the granularity of per core and per LLC way. 
Based on this technique, recent works~\cite{selfa2017application,196286,el2018kpart,xu2018dcat,xiang2018dcaps,park2019copart,rldrm} have developed approaches to dynamically allocate cache resources for applications with different characteristics, and thus achieve higher performance and resource utilization. 
However, all of them only consider the cache interference from the core side, but not the I/O side.
CacheDirector~\cite{farshin2019make} proposes a means to better utilize DDIO feature by directing the most critical data to the core's local LLC slice, but it does not consider the performance interference. 
In this paper, we demonstrate that high-speed I/O in modern servers can also introduce interference to CPU's LLC and proposes \arch to mitigate the issue. 
\arch is complementary to these prior works and can work with them to provide more comprehensive LLC management solutions.

\subsection{I/O Performance Partitioning}
There are plenty of works related to partitioning I/O for different applications/tenants. 
For example, QJUMP~\cite{188942} gives applications different priorities to reduce network interference in the data center.
Loom~\cite{227655} leverages packet scheduling in NIC to achieve performance isolation.
mClock~\cite{10.5555/1924943.1924974} developed an algorithm for I/O resource allocation in a hypervisor.
IOFlow~\cite{thereska2013ioflow}, VDC~\cite{186169}, and GIFT~\cite{246178} propose system-level I/O bandwidth isolation solutions for the storage service in data centers. 
PARTIES~\cite{Ferdman:2012:CCS:2150976.2150982} considers both I/O and CPU resources for collocated latency-sensitive tenants.
Snap~\cite{10.1145/3341301.3359657} and TAS~\cite{10.1145/3302424.3303985} deal with multi-tenancy and high data rate I/O by encapsulating the network in a single service, lowering tenant/application management complexity.
While these solutions provide isolation from different levels (device, OS, application, \etc), none of them has investigated the interference of I/O to CPU's LLC, which inevitably leads to applications' performance drop in I/O intensive scenarios. 
\arch provides the capability of identifying and alleviate such interference and thus can work with those I/O partitioning techniques together for better end-to-end performance isolation.

%% file: conclusion.tex
\section{Conclusion}
\label{sec:concl}
In modern cloud servers with Intel DDIO technology, I/O has become an important factor that affects the performance and utilization of CPU's LLC. 
In this paper, we first summarized two problems caused by DDIO, and then proposed \arch, the first I/O-aware mechanism for LLC management, which allocates LLC ways for not only the core but also the I/O. 
Our experiments showed that \arch is able to effectively reduce the performance interference caused by DDIO between applications. 
We hope this paper can attract more attention to the study of I/O-aware LLC management in the system and architecture communities.